\documentclass[12pt]{iopart}

\usepackage{graphicx}
\usepackage{dcolumn}
\usepackage{bm}
\usepackage{caption}
\usepackage{subfloat}
\usepackage{subfig}
\usepackage{float}
\usepackage{color}
\usepackage{setspace}
\usepackage{amssymb}
\usepackage{color,soul}
\usepackage{braket}

\begin{document}

 \title{Topos Many-Node Theory: Roots, Foundations and Predictions}

\author{Hamidreza Simchi}
\ead{simchi@alumni.iust.ac.ir}
\address{Department of Physics, Iran University of Science and Technology, Narmak, Tehran 16844, Iran}

\date{\today}

\noindent Among the various existing theories, we show how the concept of the space-time network has entered the physics of quantum gravity by reviewing the theories of loop quantum gravity and causal sets. Assuming that the first creatures of creation create a network, it is shown that how the network can be mapped to a topos discrete quantum manifold which is equipped with both the discrete calculus and the Alexandrov's algebra. We assign a locale to each nodes of the space-time network and show that in general, invariance under Lorentz transformations is no longer true. it is shown that the cosmological constant is non-zero and is proportional to the second power of the Hubble radius. By considering a population (set), including newly born timid children and non-timid children who survive until the birth of the new network, it is shown that the entropy of the space-time network is quantized and increases as the network grows. In consequence the inflation of the world is expected phenomenon. Also, we show that how world inflation can be described based on the concept of truth object and truth value belong to the topos theory. Although, the temperature and pressure are both high in the early moments of the creation of the world, it is shown that the quanta of vibrations, called netons, can be attributed to the vibration of space-time network, and it is expected that they will be observed in the future experiments related to cosmic background radiation. Finally, it is shown that the root of noncommutative geometry is in attributing the locale to the nodes of the space-time network instead of a point. This theory, which is quantum-relativistic from the beginning, has not the problem of a point particle, the concept of probability is an emergent concept, it does not include the problem of the measurement theory, it is not a universal Lorentz invariant, its cosmological constant is non-zero and is proportional to the second power of the Hubble radius, it describes the inflation of the universe and it has a non-commutative geometry, is called the many-node theory.

\section{Introduction}
\noindent The interaction between objects based on Newton's law of gravity, the interaction between electric charges, the interaction between electric charges and the magnetic field, and electromagnetic induction based on Faraday's law all indicate the fact that every entity has an existential effect. Usually, in physics, we define existence with the concept of particle and the essence of existence with the concept of field, and these existence effects (field) can be identified by using the appropriate existence test. But finally, we believe that there are two entities in the universe called the particle and the field, which of course we seek to turn this duality into a unity by developing the theory of quantum gravity. In models based on quantum gravity network, we show the entities with nodes and the connection and interaction between them with links. In structural formalism and structuralism logic, originality can be attributed to the network (links), to entities (nodes) or both [1]. Our assumption here is that both existence and existential originality have originality. 

Although the introduction of the space-time field based on Einstein's theory of relativity was helpful in understanding the gravitational interaction between massive bodies, it presented us with three serious challenges. First, we have to prove the existence of this space-time field experimentally [2]. Secondly, we must explain how this field interacts with other fields such as Klein-Gordon, Maxwell, and Dirac and present how to prove it experimentally. Thirdly, we must show how the quantum theory of this field is in very small dimensions, i.e. in the Planck scale, and what physical quantities we should expect to observe in the laboratory at this scale.

Loop quantum gravity theory is one of those theories that have been developed by people to answer the aforementioned challenges. In this theory, Ashtekar's variables [3, 4] are used to write the Lagrangian and then the Hamiltonian. The space-time has a $\mathrm{\Sigma }\ \times R$ topology [5] and instead of using the connection representation [6, 7], the loop representation is used to define the wave function [8-12]. there are $(N+1)\times (N+1)$ matrices with $N=1,2,3,\dots $ , belong to the different representations, which can be used as connection in loop representation. It means that the parallel transport operator along an open curve is a matrix where the indices of those matrices are tied up at intersections of curves by intertwiners. The resulting object is called spin network [4, 13-15].

The causal set theory is another theory which has been developed by people to respond to the challenges facing the space-time field and leads to the creation of the concept of the space-time field network [16-26]. Although, a continuum geometry ought to be invariant under local scale transformation [16], but, at least on superficial examination, the theory is inconsistent with experimental facts. For example, it is expected that the intrinsic scale of an atom does not depends on the history of that atom. It can be shown that the conventional positive definite metric topology of the real four dimensional space-time continuum of special relativity can be replaced by a new fine topology which is related to the causal structure [17, 18]. It means that the causal structure of a space-time, together with a conformal factor, determine the metric of a Lorentzian space-time uniquely i.e., by using the before and after relations amongst all events one can recover the conformal metric [17, 18]. Also, one can recover the entire metric and space-time if he/she has a measure for the conformal factor [17, 18]. In consequence, the metric should be defined without using the concept of the distance between neighboring points.  Instead, the concept of absolute time ordering, or causal ordering, of space-time points, events has been introduced as the one and only fundamental concept of a discrete space-time geometry [19]. Also, by considering oscillations about a flat background metric, one can solve the complete classical Euler Lagrange equations for a scattering process perturbatively [20]. But in continuum space-time, he/she encounters large wave umbers and infinitely large momenta which give rise to divergences [20]. Since, the distance between two very close points is no longer a well-defined concept, the metric is no longer well-defined [20]. Therefore, for solving the difficulty, the fundamental structure should be a discrete set of points, endowed just with a dynamical determined causal structure, such that an element is greater than another element if it lies to the future of the later (i.e., causally influenced). Such structure is called a causal set [20].  It has been shown that a Lorentzian manifold can approximate a causal set [21, 22]. A causal set (causet) can be represented graphically by a Hasse diagram [23]. A family is constructed by adjoining a single maximal element to a given causet.  Each one is a child of the parent. If we only consider the timid child (the future of every element of the parent) in each step of growth of the diagram and do not consider the gregarious child (spacelike to every other element) [23, 24] a complex network is constructed by the causet [25, 26].

Based on the results of loop quantum gravity theory and causal set theory, we can assume a fundamental principle that in the first moments of the creation of the primitive beings of the world, they create a network, where the nodes of the network stand for the entities and the links of the network stand for their causal connections and interactions. Now, the question that can be raised is: whether the nodes of this network can be mapped to the points of a discrete manifold, on which, a discrete calculus can be defined and has two topologies (one for classical gravity and the other for non-classical quantum physics) which can be defined based on the topos theory?

People have shown that by using the concept of geometric embedding [27], each node of the network can be mapped to a vector in $n$-dimensional space. On the other hand, they have shown that in a $n$-dimensional differentiable manifold, by using a coordinate function, which is a map from the open sets covering the manifold to $R^n$, a vector in $R^n$ can be attributed to each point of the manifold [28]. In this case, it can be imagined that the network nodes are mapped to a manifold so that a vector is assigned to them. Now, the question that can be raised is: whether it is possible to use a discrete manifold with classical-quantum double topology based on topos theory, which is equipped with discrete calculus, instead of the continuous $n$-dimensional differentiable classical manifold?

People have shown that a discrete set $M$  equipped with a differential calculus may regarded as a kind of analogue of a (continuous) differential manifold [29].That is, a discrete differential manifold is a discrete set $M$ together with a differential calculus on it [30, 31]. For example, if $M$ be a subset of $Z^n$ and $\mathrm{\Omega }\left(M\right)={\oplus }^{\infty }_{r=0}{\mathrm{\Omega }}^r(M)$ be a $Z$-graded associative algebra which ${\mathrm{\Omega }}^0\left(M\right)$ is the $C$-valued functions on $M$, it can be shown that $(Z,\mathrm{\Omega }\left(Z\right))$ is a mathematical model for parameter space of discrete time [30, 32]. Also, in the past, people have studied the concept of quantum topology and tried to define its features to present the theory of quantum gravity [33-36]. But, at the heart of general relativity, the manifold $M$ is locally homeomorphic to $R^n$, and in quantum mechanics, observables become operators on the Schwartz space $S(R^n)$ of fast-decreasing functions over $R^n$ [37]. Therefore, a quantum manifold ${(M}_Q)$ should be, on the one hand, locally homeomorphic to the $S(R^n)$, and, on the other hand, allows the computation of position expectation values, that recover the classical manifold [37]. It means that, there are two kinds of topology which are the expectation value topology $\iota (S^{\neq 0}\left(R^n\right),\overline{\boldsymbol{Q}}$) where $\overline{\boldsymbol{Q}}$ is the position expectation value map $\overline{\boldsymbol{Q}}:\ S^{\neq 0}(R^n)\to R^n$ and standard topology on $R^n$ $(M_Q\to S^{\neq 0}(R^n)\to R^n\leftarrow M\leftarrow M_Q)$ [37]. That is, in contrast to the usual definition of an atlas, two different topologies are introduced and $M_Q$ is a differential infinite dimensional manifold locally homeomorphic to $S^{\neq 0}(R^n)$ [37]. But, what should be a discrete set $M_Q$ and a differential calculus on it for defining the discrete quantum manifold, $M_Q$? 

In this article, we seek to introduce a theory that is primarily quantum relativistic and can provide answers to the problems related to quantum gravity, such as point particles, non-invariance under Lorentz transformations, inflation of universe, and non-commutative geometry. By reviewing the loop quantum gravity theory and the causal sets theory, we show that how the concept of the space-time network entered the physics related to the theory of quantum gravity. Then, by defining a topos discrete quantum manifold, it is shown that how we can equip this manifold with discrete calculus and Alexandrov's algebra and map the space-time network to it. By defining the concept of locale, instead of assigning a point to each node of the space-time network, we show that Lorentzian invariance is no longer valid in general, and the concept of locale and its inclusion in the theory of quantum gravity is the root of noncommutative geometry. We also show that how the inflation of the world can be explained based on both the concept of locale and the population (set) consisting of timid and non-timid children. By assuming the vibrations of space-time network, it is shown that we should expect to see quanta of this network, called netons, in the future cosmic background radiation experiments. It should be noted that at the beginning of each section of this article, first some issues published in this field are reviewed and then the issues related to this article are presented so that the roots related to the subject are discussed first and then the issues related to this article are presented, although this makes the article a little long.

The structure of article is as follows. The network of space time and topos discrete quantum manifold are provided in section 2 and 3, respectively.  Lorentzian invariance and cosmological constant are discussed in section 4 and 5, respectively. The problem of inflation of universe is justified in section 6. The vibration of the space-time network is studied in section 7 and the noncommutative geometry is discussed in section 8. The conclusion is provided in section 9.

\section{Network of space-time}

\noindent According to the experience we have with the Maxwell field and its generalized type, the Yang-Mills field, it seems that if we can formulate the space-time field in a canonical way, then we will be able to understand its interaction with other fields and its behavior in Planck-scale. Ashtekar has considered the $SU(2)$ Yang-Mills connection $A^i_a$ ($i,a=1,2,3$) ( similar to the vector potential in Maxwell theory) and constituted the configuration variables such that the densitized triads ${\tilde{E}}^a_i$ ( similar to the electric field in Maxwell theory) being their canonically conjugated momentum [3, 4]. By assuming that the space-time has a $\mathrm{\Sigma }\ \times R$ topology and that there is a time-like direction characterized by a vector $t^{\mu }$ whose orbits (curves such that their tangents are the vector of interest) are a curve parametrized by a parameter $t$ and such that $t=$ constant surface are spatial slices $\mathrm{\Sigma }$, we can decompose $t^a=Nn^a+N^a,\ a=1,2,3$, where $n^a$ is in-plane of $\mathrm{\Sigma }$ and $N^a$ is normal to $\mathrm{\Sigma }$ .The quantities $N$ and $N^a$ are usually known as the lapse function and shift vector, respectively [4, 5]. By using $\tilde{N}$  (lapse function with density weight $-1$), $N^a$, $A^i_a$, and ${\tilde{E}}^a_i$, one is able to write the Lagrangian of space-time field [3, 4]. Of course, by considering the gauge transformation of Yang-Mills theory, it can be shown that the term ${\lambda }^i{(D_a{\tilde{E}}^a)}^i$ should be added to the Lagrangian where ${\lambda }^i\ $are Lagrange multipliers and $D_a\ $is covariant derivative [3, 4].  The theory has nine configuration degree of freedom in the $A^i_a$'s and seven constraints, which leaves space-time filed as a theory with two degree of freedom, just like Maxwell theory. Now, one is able to write the Hamiltonian of space-time filed as a combination of constraints times Lagrange multipliers and find its interaction with the other fields [4, 5-7].  In connection representation, we can consider ${\hat{A}}^i_a\to A^i_a$ as a multiplicative operator, ${\widehat{\tilde{E}}}^a_i\to -i\frac{\delta }{\delta A^i_a}$ as functional derivatives, and $\mathrm{\Psi }(A^i_a)$ as wave function [4].  Now by using  ${\hat{A}}^i_a$,  ${\widehat{\tilde{E}}}^a_i$, and  $\mathrm{\Psi }(A^i_a)$,  one is able to find the communication relations, quantized equation of constraints and gauge transformation [4]. Of course, there are some difficulties (not know how to work with $\mathrm{\Psi }(A)$ in a controlled mathematical fashion, suitable inner product, and the difficulties of promoting the Hamiltonian constraint to an operator) in connection representation and in consequence people have tried to develop an alternative representation called loop representation [4, 8-11].  

The foundations of the loop representation are based on the concept of holonomy and Giles' theorem. Due to the Stokes's theorem, the circulation of the vector potential along a closed curve is equal to the surface integral of its curl. The curl of the vector potential is important since it is proportional to the filed tensor in Maxwell theory. Therefore, if one specifies the circulation of the vector potential for all possible curves, he/she is implicitly specifying the field. In Yang-Mills theories, we encounter two problems. First, the field tensor is something more than just the curl of a vector. Second, the circulation of the vector potential is not gauge invariant [4]. The concept of holonomy is introduced to play the role of circulation of the vector potential in Yang-Mills theories [4]. In the other words, we take a quantity such as triads $E^a_i$ and we want to carry it around a curve ${\gamma }^a\left(t\right)$ in space as parallel to itself as possible. It can be done by using the parallel propagator, since it is an operator that takes from $0$ to $t$ as parallel as possible [4]. If ${\gamma }^a\left(t\right)$ coincides with ${\gamma }^a(0)$ so one is propagating along a closed curve, then such a parallel propagator is called holonomy which is a matrix and its trace is a scalar. Since, the scalar is invariant under gauge transformation, it can be a good candidate for an observable quantity for the Yang-Mills theories [4]. It should be noted that unlike in the Maxwell theory, the Poisson bracket of Gauss' law with the field tensor is non-vanishing and the field tensor transforms under gauge transformation in Yang-Mills theories [4]. Therefore, the electric and magnetic field are not observable quantities in Yang-Mills theories since they are gauge dependent [4]. The Giles' theorem [12] states that if you know the trace of holonomy along all possible loops in a manifold for a given vector potential you can reconstruct from their values all the gauge invariant information present in the vector potential. Therefore, the trace of holonomy matrix is a basis for all possible observables that are function of the connection only.

Using the above aforementioned concepts, a state in loop representation is written as $\mathrm{\Psi }\left(A\right)=\sum_{\gamma }{\mathrm{\Psi }\left[\gamma \right]W_{\gamma }[A]}$ where the $\sum_{\gamma }{(\cdots )}$ is a formal sum over all possible loops, $\mathrm{\Psi }\left[\gamma \right]$ are functions that depend on a loop, and the trace of holonomies $W_{\gamma }[A]$ is equal to $Tr\left(P\left[exp\left(-\oint^{\gamma (t)}_{\gamma (0)}{{\dot{\gamma }}^a\left(s\right){\boldsymbol{A}}_a\left(s\right)ds}\right)\right]\right)$. Here, $Tr$ and $P$ stand for trace and path ordering, $\dot{\gamma }\left(t\right)=d{\gamma }^a(t)/dt$ is the tangent vector to the curve and ${\boldsymbol{A}}_a$ is connection vector. The connection that one uses to describe space-time filed when using Ashtekar's variables is $su(2)$ connection with Levi-Civita symbols as structure constants [4]. Of course, there are $(N+1)\times (N+1)$ matrices with $N=1,2,3,\dots $ , belong to the different representations, which can be used as connection. Therefore, one can use matrices in any representation to construct a connection and, with it, parallel transport operators along curves. It means that the parallel transport operator along an open curve is a matrix where the indices of those matrices are tied up at intersections of curves by intertwiners [4, 13-15]. The resulting object is called spin network. It is a graph with intersections such that  a number $N$ (or $N/2$) is assigned to each line of the graph. The number $N$ (or $N/2$) is associated with the dimensionality of the matrices of the holonomies [4, 13-15]. 

Another theory that has been developed by people to respond to the challenges facing the space-scale field and leads to the creation of the concept of the space-time field network is the causal set theory. Non-experimentally verified continuity of space-time has survived both the revolutions of relativity and quantum mechanics. It is expected that the discreteness of space-time ought to lead to observable effects and pays the way for avoiding all the conceptual difficulties of the non-countable infinite continuum and the divergence problems connected with the concept of a point particle. Based on the Weyl's theory [16], a continuum geometry ought to be invariant under local scale transformation.

Let us, consider the real four dimensional space-time continuum of special relativity ($M$) and its characteristic quadratic form (Q). If on $M$, $y-x$ is a time vector ($x<y$), then $Q(x-y)>0$, oriented toward the future, $x_0<y_0$. By considering a one-to-one (linear or continuous) mapping $f:M\to M$, such that $f$ and $f^{-1}$ preserve the partial ordering i.e., $x<y\Leftrightarrow fx<fy$ for all $x,y\in M$, we call $f$ a causal automorphism. The causal automorphisms form a group which is called causality group [17]. Let $G$ be the group generated by (i) linear maps of $M$ that leave $Q$ invariant, and preserve time orientation, but possibly reverse space orientation (orthochronous Lorenz group), (ii) translation of $M$ and (iii) dilatations of $M$ (multiplication by a scalar) then the causality group is $G$ [17]. It can be shown that the conventional positive definite metric topology of $M$ can be replaced by a new fine topology which is related to the causal structure [17]. However, the topology has some disadvantages which are (i) a three-dimensional section of simultaneity has no meaning in terms of physically possible experiments, (ii) the homothecy group of $M$ is not significant physical, (iii) The set of the topology-group paths does not incorporate accelerating particles moving under forces in curved lines, and (iv) the topology is technically complicated [18]. It is possible one propose a new topology for strongly causal space-time M which share the attractive features of the previous topology, but which also answer the aforementioned disadvantages and have additional attractive physical features [18]. The new topology (path topology) (i) is defined to the finest topology on $M$ which induces the Euclidean topology on arbitrary timelike curves, (ii) incorporates the causal, differential, and smooth conformal structure, (iii) incorporates all timelike paths, and (iv) is still technically complicated, but less so than the previous one [18]. By considering the new topology, it can be shown that the causal structure of a space-time, together with a conformal factor, determine the metric of a Lorentzian space-time uniquely i.e., by using the before and after relations among all events, one can recover the conformal metric [17, 18]. Also, one can recover the entire metric and space-time if he/she has a measure for the conformal factor [17, 18].  

The discreteness of space-time can be concluded from other different points of view, too. As we mentioned before, a continuum geometry ought to be invariant under local scale transformation [16]. But, at least on superficial examination, the theory is inconsistent with experimental facts. For example, it is expected that the intrinsic scale of an atom does not depends on the history of that atom [19]. Since, the local scale of the metric is not arbitrary, the space-time should be discrete. Otherwise one needs something like a fundamental mass scale for fixing the scale of the metric. Then the origin of this mass scale would have to be looked for outside of geometry [19]. It should be noted that defining the theory in terms of the distance between neighboring points could be done without referring to a background geometry. But it would nevertheless be a rather artificial procedure and in consequence the metric should be defined without using the concept of the distance between neighboring points.  Instead, the concept of absolute time ordering, or causal ordering of space-time points (events) has been introduced as the one and only fundamental concept of a discrete space-time geometry [19]. Therefore, space-time is nothing but the causal ordering of events [17]. In consequence, the counting measure is a natural measure, and the causal ordering alone is the only structure needed and the coordinates and metric may be derived as secondary concepts [19].  

Now let us,  consider oscillations about a flat background metric with signature $(+,+,+,+)$ as a small perturbation i.e., $g_{\mu v}={\delta }_{\mu v}+h_{\mu v}$  and the Lagrangian as $L=-\sqrt{g}R$ where $R$ is curvature and now $g>0$ [20].  One  can expand $L$ in terms of $h_{\mu v}$ up to the quadratic terms in $h_{\mu v}$.  The linear parts of the wave equation can be quantized in energy unit $\hslash v$, called gravitons, and higher order terms implies that the equations for gravitational waves are non-linear so that one can get scattering i.e., the higher order terms in the Lagrangian now cause gravitons to interact [20].  But, there is an invariance in the system under infinitesimal coordinate transformations. Therefore, if $C_{\mu }=h_{\mu \alpha ,\alpha }-\frac{1}{2}h_{\alpha \alpha ,\mu }$, and $L=-\sqrt{g}R-\frac{1}{2}C^2_{\mu }$, then $L=-\frac{1}{4}{\left(h_{\alpha \beta ,v}\right)}^2+\frac{1}{8}$${\left(h_{\alpha \alpha ,v}\right)}^2+O\left(h^3\right)+\cdots $. Now, if one has the inverse of the Laplace operator, called propagator for the graviton, and the explicit form of the interaction terms $O\left(h^3\right)$ , called vertices, he/she is able to solve the complete classical Euler Lagrange equations for a scattering process perturbatively [20].   Nevertheless, the theory dictates that gravitons can split, from closed loops and rejoin while they scatter. Therefore, if one computes the contributions coming from such diagrams one find that they are actually highly divergent and metric tensor no longer makes sense [20]. They mean that there may be something basically wrong with working in a continuous space-time, because that is where all the difficulties came from.  In continuum space-time, we encounter large wave umbers and infinitely large momenta which give rise to divergences [20]. Also, the metric is no longer well-defined if one wants to measure distance between two very close points because this distance is no longer a well-defined concept [20]. Therefore, similar to the previous case [19], the fundamental structure is a discrete set of points, endowed just with a dynamical determined causal structure i.e., we consider a set in the form of a partial order relation, such that an element is greater than another element if it lies to the future of the later (i.e., causally influenced). Such structure is called a causal set [20].  Now, one question can be asked: if the space-time is a causal set how can we relate it to the picture of space-time as a continuum? If we consider an arbitrary causal set containing many elements, which manifold with metric it looks like at large scale? It has been shown that a Lorentzian manifold can approximate a causal set, noting in particular that the thereby defined effective dimensionality of a given causal set can vary with length scale [21, 22].

A causal set (causet), which is a partially ordered set (or poset) can be represented graphically by a Hasse diagram [23]. An element of the poset is shown by a dot and the relation $x<y$   between two  elements is shown by a link such that the preceding element $x$ is drawn below the following element $y$ [23].  A family is constructed by adjoining a single maximal element to a given causet.  Each one is a child of the parent. If the adjoined element is to the future of every element of the parent, it is called timid child and if it is spacelike to every other element it is called gregarious child [23, 24]. If we only consider the timid child in each step of growth a complex network is constructed by causet [25, 26].

Based on the above descriptions, we can consider a given Hasse diagram as a parent graph and adjoin a timid child (or event) to each node of the graph and consider connection and interaction between nodes as links and call the resulted graph a complex network of space-time field.

\section{Topos Discrete Quantum Manifold}

Given a finite, without loops and multiple edges graph $G$, a representation of $G$ is a map that assigns to each vertex $x$ of $G$ a vector $\overrightarrow{x}\in R^d$ such that two vertices $x$ and $y$ are joined in $G$ if and only if $\overrightarrow{x}$ and $\overrightarrow{y}$ satisfy some specified geometrical conditions e.g., $\overrightarrow{x}\bullet \overrightarrow{y}\ge t\in R$ where $\overrightarrow{x}\bullet \overrightarrow{y}$  is the scalar product and $t$ is a real threshold value, admitting also negative values [27]. When $t=1$, the representation of $G$ is the simplest one and the scalar product dimension $d(G)$ of a graph $G$ is defined to be the minimum number $d\ge 1$ such that $G$ admits a representation in $R^d$ [27]. In the simplest representation, if all vectors $\overrightarrow{x}$ representing vertices $x$ are unit i.e., $\left\|\overrightarrow{x}\right\|=1$, the representation is called spherical and the minimum number $d$ is called the spherical dimension of $G$ and denoted by $sd(G)$ [27]. If $d$ is minimum and for every $x$ and $y\ $belong to $G$, then $\left\|\overrightarrow{x}-\overrightarrow{y}\right\|\le \rho \in R$, the representation is called a distance representation of $G$ and minimum number $d$ is called the sphericity and denoted by $sph(G)$ [27]. It can be shown that $(sd\left(G\right)-1)\le sph(G)\le sd(G)$ and $sd(G)\ge d(G)$, in general [27]. The maximum distance of a vertex to all other vertices is considered as the eccentricity of graph and denoted by $e(V)$. The minimum among all the maximum distance between a vertex to all other vertices is considered as the radius of the graph $G$ and denoted by $r(G)$. It can be found by finding the minimum value of $e(V)$ from all the vertices. The size of graph is its number of edges $\left|E\right|$. However, in some context, the size is $\left|V\right|+\left|E\right|$ where $\left|V\right|$ is its number of vertices. For a given graph $G=(V,E)$, an independent set $S$ is a maximal independent set if for $v\in V$, then $v\in S$ or $N(v)\cap S\neq \emptyset $ where $N(v)$ denotes the neighbors of $v$. It can be shown that
\\
\begin{equation}
sd(G)\ge sph(G)\ge \frac{Ln\ (\alpha \left(G\right))}{Ln(2r\left(G\right)+1)}
\end{equation}

and

\begin{equation}
sd(G)\le 16{\left(d+1\right)}^3Ln(8n\left(d+1\right))
\end{equation}

where $r(G)$, $\alpha (G)$, $d$, and $n$ stand for radius, a maximal independent set of size, maximum degree, and number of vertices of graph, respectively [27].

It is well known that if $M$ is a topological space which is provided with a family of pairs $\{\left(U_i,{\varphi }_i\right)\}$ where $\{U_i\}$ is a family of open sets such that ${\cup }_iU_i=M$ and ${\varphi }_i$ is a homeomorphism from $U_i$ onto an open subset $V_i$ of $R^n$ such that for given $U_i\cap U_j\neq \emptyset $ the map ${\psi }_{ij}={\varphi }_i\circ {\varphi }^{-1}_j$ from ${\varphi }_j(U_i\cap U_j)$ to ${\varphi }_i(U_i\cap U_j)$ is infinitely differentiable, $M$ is an $n$-dimensional differentiable manifold [28]. The pair $\left(U_i,{\varphi }_i\right)$ is called a chart and the whole family $\{\left(U_i,{\varphi }_i\right)\}$ is called atlas. $U_i$ is called the coordinate neighborhood while  ${\varphi }_i$ is the coordinate function. The set $\left\{x^{\mu }\left(p\right)\right\}=\{x^1\left(p\right),\cdots ,x^n\left(p\right)\}$ is called coordinate [28]. Therefore, it seems natural that by mapping the vertices of a graph to a discrete manifold, the coordinates of that manifold are mapped to the vertices, and in fact, the same act of embedding occurs.

Let us, consider a countable set $M$ with elements $i,j,\cdots $ and the $C$-valued functions on $M$ denoted by $A$. If the map $e:M\to M$ such that $e\left(i\right)=i$ i.e., $e_i\left(j\right)={\delta }_{ij}$ then $\sum_i{e_i\left(j\right)=\sum_i{{\delta }_{ij}}}$ $\to $ $\sum_i{e_i=\mathrm{l}}$ where $\mathrm{l}\left(i\right)=1$ [29]. Now, each map $f\in A\ (f:M\to M)$ can be written as $f=\sum_i{f(i)e_i}$ [29]. It is possible one extends the algebra $A$ to a differential algebra $(\mathrm{\Omega }\left(M\right),d)$ such that $\mathrm{\Omega }\left(M\right)={\oplus }^{\infty }_{r=0}{\mathrm{\Omega }}^r(M)$ be a $Z$-graded associative algebra where ${\mathrm{\Omega }}^0\left(M\right)=A$ and $d:{\mathrm{\Omega }}^r(M)\to {\mathrm{\Omega }}^{r+1}(M)$ is a linear operator which satisfies $d^2=0$, $d\mathrm{l}=0$ , and for ${\omega }_1,{\omega }_2\in {\mathrm{\Omega }}^r(M)$ then (Leinbiz rule) [29]

\begin{equation}
d\left({\omega }_1{\omega }_2\right)=d\left({\omega }_1\right){\omega }_2+{(-1)}^r{\omega }_1d{\omega }_2 
\end{equation}

Therefore
\begin{equation}
 de_i=\sum_j{(e_{ji}-e_{ij})}, df=\sum_{i,j}{e_{ij}(f\left(j\right)-f\left(i\right))} 
\end{equation}
\begin{equation}
de_{ij}=de_ide_j=\sum_k{(e_{kij}-e_{ikj}+e_{ijk})}
\end{equation}

and any 1-from can be written as [29] $\rho =\sum{e_{ij}{\rho }_{ij}}$ with ${\rho }_{ij}\in C$ and ${\rho }_{ii}=0$ with

\begin{equation}
d\rho =\sum_{i,j,k}{e_{ijk}({\rho }_{jk}-{\rho }_{ik}+{\rho }_{ij})}
\end{equation}

 Finally [29], any $\psi \in {\mathrm{\Omega }}^{r-1}(A)$ can be written as $\psi =\sum_{i_1\cdots i_r}{e_{i_1\cdots i_r}{\psi }_{i_1\cdots i_r}}$ with ${\psi }_{i_1\cdots i_r}\in C$, and if $i_s=i_{s+1}$ for some $s$ then ${\psi }_{i_1\cdots i_r}=0$ and

\begin{equation}
d\psi =\sum_{i_1\cdots i_{r+1}}{e_{i_1\cdots i_{r+1}}\sum^{r+1}_{k=1}{{(-1)}^{k+1}{\psi }_{i_1\cdots {\hat{i}}_k\cdots i_{r+1}}}}
\end{equation}

 It is called universal differential algebra (envelope of $A$) if no further relations are imposed on the differential algebra [29]. If we are interested to approximate a differential manifold by a discrete set, the universal differential algebra is too large to provide us with a corresponding analog of the algebra of differential forms on the manifold. Therefore, to each element of $M$ a vertex of a graph, and to each $e_{ij}\neq 0$, an arrow from $i$ to $j$ are assigned. Now, the universal differential algebra corresponds to the graph where all the vertices are connected pairwise by arrows in both directions. Deleting arrows leads to a graph which represents a reduction of the universal differential algebra [29].

For example, a field $\psi $ on $M$ is a cross section of a vector bundle over $M$ and $G=\sum_i{G(i)e_i}$ are elements of a subgroup of $GL(n,A)$ on nontrivial bundles correspond to finite projection modules, denoted by the free $A$-module $A^n$, and stand for an action $\psi \to G\psi $ of a local gauge group. Then, it induces on the dual (right $A$-) module an action $\alpha \to \alpha G^{-1}$. Now, the covariant exterior derivatives are defined as $D\psi =d\psi +A\psi $ and $D\alpha =d\alpha -\alpha A$, where $A$ is a 1-form [29]. It can be shown that if $\boldsymbol{A}$ obeys the usual transformation law of a connection 1-form then a transport operator can be written as $U=\sum_{i,j}{e_{ij}}{(\boldsymbol{1}+A_{ij})}$ such that an arrow from $j$ to $i$ is assigned to $U_{ij}$. The curvature of the connection \textbf{A} can be written as
\begin{equation}
 F=\sum_{i,j,k}{e_{i,j,k}(U_{ij}U_{jk}-U_{ik})}
\end{equation}

if we consider the covariant derivatives as [29]
\begin{equation}
 D\psi =\sum_{i,j}{e_{ij}(U_{ij}\psi \left(j\right)-\psi \left(i\right))} 
\end{equation}
and 
\begin{equation}
D\alpha =\sum_{i,j}{e_{ij}(\alpha \left(j\right)U^{-1}_{ij}-\alpha \left(i\right))}
\end{equation}
 where $\psi =\sum_i{e_i\psi (i)}\ $[29]. Based on the universal differential algebra, the Yang-Mills action becomes [29]
\begin{equation}
S_{YM}=tr\sum_{i,j,k}{(U^{\dagger }_{jk}}U^{\dagger }_{ij}U_{ij}U_{jk}-U^{\dagger }_{jk}U^{\dagger }_{ij}U_{ik}-U^{\dagger }_{ik}U_{ij}U_{jk}+U^{\dagger }_{ik}U_{ik}) 
\end{equation}

with $U^{\dagger }_{ij}=U_{ji}$ and in consequence the Yang-Mills equation is
\begin{equation}
\sum_k{\left(F_{ikj}-{\delta }_{ij}F_{iki}-U_{ik}F_{kij}-F_{ijk}U_{kj}\right)=0}
\end{equation}

Therefore, the discrete set and differential calculus can be used for studying the Yang-Mills theories such as gravity in its covariant forms. 

A discrete set $M$  supplied with a differential calculus may regarded as a kind of analogue of a (continuous) differential manifold [29]. That is, a discrete differential manifold is a discrete set $M$ together with a differential calculus on it [30, 31]. For example, let $M$ be a subset of $Z^n$, then
\begin{equation}
x^{\mu }=\sum_{a\in M}{a^{\mu }}{e_{a}}, (\mu =1,\cdots ,n)
\end{equation}

are natural coordinate functions on $M$. Since, $e_ide_j=-de_ie_j+{\delta }_{ij}de_j$, $\sum_i{de_i=0}$, and, $de_{i_1\cdots i_r}=\sum_j{\sum^{r+1}_{k=1}{{(-1)}^{k+1}e_{i_1\cdots i_{k-1}ji_k\cdots i_r}}}$, one obtains [30]
\begin{equation}
\left[dx^{\mu },x^v\right]=\sum_{a,b}{(a^{\mu }-b^{\mu })(a^v-b^v)\equiv {\tau }^{\mu v}}
\end{equation}
and
\begin{equation}
\left[{\tau }^{\mu v},x^{\lambda }\right]=\sum_{a,b}{(a^{\mu }-b^{\mu })(a^v-b^v)(a^{\lambda }-b^{\lambda })e_{ab}}
\end{equation}
If, the differential calculus (called the oriented lattice calculus) is determined by $e_{ab}\neq 0\Longleftrightarrow b=a+\widehat{\mu }$ where $\widehat{\mu }=({\widehat{\mu }}^{v})=({\delta }^{v}_{\mu })$ then ${\tau }^{\mu v}={\delta }^{\mu v}\sum_a{e_{a,a+\widehat{\mu }}={\delta }^{\mu v}dx^{\mu }}$ and $\left[dx^{\mu },x^v\right]={\delta }^{\mu v}dx^v$. Now, for special case ($n=1$) which corresponds to the reduction $e_{ij}\neq 0\Longleftrightarrow j=i+1$ under condition $\left[dt,t\right]=dt$, it assigns a 1-dimensional structure to $Z$ such that $df=dt\left({\partial }_+f\right)\left(t\right)=\left({\partial }_-f\right)\left(t\right)dt$ define functions ${\partial }_{\pm }f$ on $M$ [30]. It can be shown that $\left({\partial }_+f\right)\left(t\right)=f\left(t+1\right)-f(t)$ and $\left({\partial }_-f\right)\left(t\right)=f\left(t\right)-f(t-1)$ i.e.,. We can use the notation $\dot{f}\left(t\right)=f\left(t+1\right)-f(t)$ instead of $\left({\partial }_+f\right)\left(t\right)$ [30].Then, $(Z,\mathrm{\Omega }\left(Z\right))$ is a mathematical model for parameter space of discrete time [30, 32]. Since, we are looking for a quantum theory for the space-time field, we need to determine whether a discrete quantum manifold is definable and, if so, what important properties it would have? In the past, people have studied the concept of quantum topology and tried to define its features to present the theory of quantum gravity [33-36]. 

At the heart of general relativity, the manifold $M$ is locally homeomorphic to $R^n$, and in quantum mechanics, observables become operators on Hilbert space, or more accurately, on the Schwartz space $S(R^n)$ of fast-decreasing functions over $R^n$, which is reobtained as the set of all possible position expectation values [37]. Now, one should provide a mathematical construction for unifying the manifold and function space ideas, which might be seen as the fundamental structures behind general relativity and quantum theory, respectively [37]. A quantum manifold ${(M}_Q)$ is, on the one hand, locally homeomorphic to the $S(R^n)$, and, on the other hand, allows the computation of position expectation values that recover the classical manifold [37]. A quantum atlas of dimension $n\in N$ on a set $M_Q$ is a collection of pairs $A=\{\left(U_i,{\phi }_i\right),\ i\in I)$, for some index set $I$, called charts, which satisfy the four conditions: (i) Each $U_i$ is a subset of $M_Q$ and the $U_i$ cover $M_Q$, (ii) Each ${\phi }_i$ is a bijection of $U_i$ onto a set ${\phi }_i\left(U_i\right)\subset S^{\neq 0}\left(R^n\right)=S\left(R^n\right)\backslash \{0\}$, (iii) for each $i,j$, the set ${\phi }_i(U_i\cap U_j)$ is open in the expectation value topology $\iota (S^{\neq 0}\left(R^n\right),\overline{\boldsymbol{Q}}$) where $\overline{\boldsymbol{Q}}$ is the position expectation value map $\overline{\boldsymbol{Q}}:\ S^{\neq 0}(R^n)\to R^n$ and is continuous with respect to the nuclear topology restricted to $S^{\neq 0}\left(R^n\right)$ and standard topology on $R^n$, and, (iv) for each $i$, $j$, the transition map ${\phi }_{ji}={\phi }_j\circ {\phi }^{-1}_i$ on the overlap of any two charts, ${\phi }_{ji}:\ {\phi }_i(U_i\cap U_j)\to {\phi }_j(U_i\cap U_j)$ is continuous in the expectation value topology and differentiable in the natural topology [37]. In contrast to the usual definition of an atlas, two different topologies are introduced and $M_Q$ is a differential infinite dimensional manifold locally homeomorphic to $S^{\neq 0}(R^n)$ $(M_Q\to S^{\neq 0}(R^n)\to R^n\leftarrow M\leftarrow M_Q)$ [37, 38]. Now, whether one can discretize $M_Q$ and introduce a discrete quantum manifold? In the other words, what should be a discrete set $M_Q$  and a differential calculus on it for defining the discrete quantum manifold, $M_Q$?

However, it is well known that, measurements are regarded as operations on physical systems and the quantum mechanics is concerned with the results of measurements. The mathematics required to describe the measurement procedures should be the mathematics of operators. Operators appear in mathematics in various form, the simplest operators are those represented by matrices. In matrix mechanics, one is often concerned with vectors and matrices with a countable infinite number of components and elements, respectively. In quantum mechanics, we concern with vectors in Hilbert space [39-41]. Therefore, we can assign a countable matrix Hilbert space, $H^n_{matrix}$, to quantum manifold $M_Q$ such that the Hermitian matrices (operators) act on it for calculating the result of measurements and their related expectation values i.e. now, we have: 
\[{Network}_{spacetime}\to M_Q\to H^n_{matrix}\to R^n\leftarrow M\leftarrow M_Q{\leftarrow Network}_{spacetime},\] 
where ${{Network}_{spacetime}\to M}_Q\to H^n_{matrix}\to R^n$ covers the quantum mechanical parts and $R^n\leftarrow M\leftarrow M_Q{\leftarrow Network}_{spacetime}$ covers the classical parts. In the above representation, in the left hand side of $R^n$ ,we deal with matrices which includes countable discrete elements and in the right hand side of $R^n$, we deal with discrete calculus. Therefore, we concern with the discrete quantum manifold, $M_Q$, which is equipped by two topologies $\iota (H^n_{matrix},R^n)$ and $\kappa (M,R^n)$. For example, in Yang-Mills theories as it was shown above, by using the differential calculus on $M$, the field tensor can be written as $F=\sum_{i,j,k}{e_{i,j,k}(U_{ij}U_{jk}-U_{ik})}$ where $U=\sum_{i,j}{e_{ij}}{(\boldsymbol{1}+A_{ij})}$ is the transport operator and $\boldsymbol{A}$ obeys the usual transformation law of a connection 1-form. Now, by using the Dirac' rules, one can find the operator (matrix) representation of $\boldsymbol{A}$\textbf{ }i.e.,\textbf{ }$\widehat{\boldsymbol{A}}$\textbf{ }and in consequence $\widehat{\boldsymbol{U}}$ and\textbf{ }$\widehat{\boldsymbol{F}}$. 

 But, it is well known that, there are some problems with quantum mechanics [42, 43]. For example, an electron spin that has not been measured is like a musical chord, formed from a superposition of two notes (spin up and down), each note with its own amplitude. In this musical analogy, the act of measuring the spin somehow shifts all the intensity of the chord to one of the notes, which we then hear on its own [42]. If we only consider the spin of electron, its wave function is just a pair of numbers, each number for each sign of its spin in some chosen direction, similar to the amplitude of each of the two notes in chord. Therefore, its wave function whose spin has not been measured generally has nonzero values for spins of both signs [42]. Due to the Born rule, the probabilities of finding either spin up or spin down are proportional to the squares of the absolute values of the complex numbers in its wave function [42]. But Schrodinger equation does not involve probabilities and it is a deterministic equation and gives the wave function at any moment [42]. Therefore, one question can be asked: how do probabilities get into quantum mechanics? [42]. One common answer is that the spin is put in an interaction with a macroscopic environment that jitters in an unpredictable way [42]. But, if the deterministic Schrodinger equation governs the changes through time, then the results of measurement should not in principle be unpredictable, of course, based on the realist approach of Bohr [42]. But, it has been shown that the existence of the superposition is undetectable, and in effect, the history of the world has split into two streams, uncorrelated with each other [44]. 

 One of the ways out of the problem related to the point particle, the presence of probability and the issue of measurement in the theory of quantum mechanics is to use the theory of topos quantum mechanics [26, 45, 46]. In this theory, the quantum theory could be seen as a collection of local classical approximation (classical snapshots) and the quantum information put back into the picture by the categorical structure of the collection of all the classical contexts [45].  In consequence, no information is lost and therefore we have no cheat [45]. If $\mathcal{B}(\mathcal{H})$ be the algebra of bounded operators on a Hilbert space ($\mathcal{H}$) and $B\subset \mathcal{B}(\mathcal{H})$ such that includes identity and is closed with respect to taking the adjoint, and the double commutant of $B$ ($C$) is equal to $B$, then $C$ is the von-Neumann algebra generated by $B$ [45]. It can be shown that, given a Hilbert space $\mathcal{H}$, the collection of all abelian von Neumann subalgebras, denoted as $V(\mathcal{H})$, forms a category [45]. $V(\mathcal{H})$ is a partially ordered set (poset), whose ordering is given by subset inclusion and can be equipped with an Alexandrov, vertical, or bucket topologies [45]. In this article, we only consider the Alexandrov algebra. It can be shown that the action of a given group $G$ on $V(\mathcal{H})$ is not continuous with respect to the Alexandrov topology while it is continuous with respect to both the vertical topology and the bucket topology [45, 47]. In topos quantum  mechanics, for each context $v\in V(\mathcal{H})$, one tries to reproduce a situation analogous to the classical physics in which self-adjoint operators are identified with functions from a space to the reals. Therefore, the topological space ${\underline{\mathrm{\Sigma }}}_v$ can be interpreted as a local state space, one for each $v\in V(\mathcal{H})$ [45]. Since, not all operators contained in a single algebra $v$, we should consider the collection of all the local state spaces [45]. A collection of local state spaces is the topos analogue of the state space which is called the spectral presheaf and denoted as $\underline{\mathrm{\Sigma }}$ [45]. In topos quantum theory, physical quantities are defined as a functor $A:\ \underline{\mathrm{\Sigma }}\to {\underline{R}}^{\leftrightarrow }$ where ${\underline{R}}^{\leftrightarrow }$ is a presheaf with objects:
\begin{equation}
{{\underline{R}}^{\leftrightarrow }}_{V}=\{(\mu ,\nu )\vert\mu ,\nu :\downarrow V\to R; \mu \le \nu \}
\end{equation}
which $\mu$ is order preserving, $\nu$ is order reversing and morphisms:
\begin{equation}
{{\underline{R}}^{\leftrightarrow }}_{V_1,V_2}:{{\underline{R}}^{\leftrightarrow }}_{V_1}\to {{\underline{R}}^{\leftrightarrow }}_{V_2}\, (\mu ,\nu )\longmapsto ({\mu \vert}_{V_2},{\nu \mathrm{\vert}}_{V_2}) 
\end{equation}
where
\begin{equation}
\downarrow W=\{v\in V(\mathcal{H})\vert v\le W\}, W\in V(\mathcal{H})
\end{equation}
 is the collection of all lower sets in $V\left(\mathcal{H}\right)$ [45]. Here, $\downarrow W$ is the basis of Alexandrov topology. This presheaf is where physical quantities take their values, thus it has the same role as the reals in classical physics [45]. Since, in quantum theory one can only give approximate values to the quantities, the quantity object is defined in terms of order reversing and order preserving values to the quantities [45]. In the other words, a physical quantity $A$ is represented by the map
\begin{equation}
\check{\delta }\left(\hat{A}\right):\underline{\mathrm{\Sigma }}\to {\underline{R}}^{\longleftrightarrow } 
\end{equation}
which, at each context $v$, is defined as
\begin{equation}
\lambda \longmapsto {\breve{\delta }}(\hat{A})_{v}=({\breve{\delta }}^{i}{(\hat{A})}_{v}(\lambda),{\breve{\delta }}^{o}{(\hat{A})}_{v}(\lambda)) 
\end{equation}
where, $\lambda $, ${\breve{\delta }}^{i}{(\hat{A})}_{v}$, and ${{\breve{\delta }}^{o}(\hat{A})}_v$ are the set of eigenvalues of the associated self-adjoint operator $\hat{A}$, order reversing function, and order preserving function, respectively [45]. In the language of canonical quantum theory, ${{\check{\delta }}^i\left(\hat{A}\right)}_v$, and ${{\check{\delta }}^o\left(\hat{A}\right)}_v$ are interpreted as the smallest and the largest results of measurements of a physical quantity $\hat{A}$ given the state  $\mathrm{\vert}$$\psi >$, respectively [45].
That is, if $\vert\psi >$ is an eigenstate of $\hat{A}$ then
\begin{equation}
\left\langle \psi \mathrel{\left|\vphantom{\psi  \hat{A} \psi }\right.\kern-\nulldelimiterspace}\hat{A}\mathrel{\left|\vphantom{\psi  \hat{A} \psi }\right.\kern-\nulldelimiterspace}\psi \right\rangle =\left\langle \psi \mathrel{\left|\vphantom{\psi  {{\delta }^o(\hat{A})}_v \psi }\right.\kern-\nulldelimiterspace}{{\delta }^o(\hat{A})}_v\mathrel{\left|\vphantom{\psi  {{\delta }^o(\hat{A})}_v \psi }\right.\kern-\nulldelimiterspace}\psi \right\rangle =\left\langle \psi \mathrel{\left|\vphantom{\psi  {{\delta }^i(\hat{A})}_v \psi }\right.\kern-\nulldelimiterspace}{{\delta }^i(\hat{A})}_v\mathrel{\left|\vphantom{\psi  {{\delta }^i(\hat{A})}_v \psi }\right.\kern-\nulldelimiterspace}\psi \right\rangle 
\end{equation}
However, if it is not an eigenvalue then
\begin{equation}
\left\langle \psi \mathrel{\left|\vphantom{\psi  {{\delta }^o(\hat{A})}_v \psi }\right.\kern-\nulldelimiterspace}{{\delta }^o(\hat{A})}_v\mathrel{\left|\vphantom{\psi  {{\delta }^o(\hat{A})}_v \psi }\right.\kern-\nulldelimiterspace}\psi \right\rangle \ge \left\langle \psi \mathrel{\left|\vphantom{\psi  \hat{A} \psi }\right.\kern-\nulldelimiterspace}\hat{A}\mathrel{\left|\vphantom{\psi  \hat{A} \psi }\right.\kern-\nulldelimiterspace}\psi \right\rangle \ge \left\langle \psi \mathrel{\left|\vphantom{\psi  {{\delta }^i(\hat{A})}_v \psi }\right.\kern-\nulldelimiterspace}{{\delta }^i(\hat{A})}_v\mathrel{\left|\vphantom{\psi  {{\delta }^i(\hat{A})}_v \psi }\right.\kern-\nulldelimiterspace}\psi \right\rangle 
\end{equation}
and 
\begin{equation}
{\breve{\delta }\left(\hat{A}\right)}_v=({{\breve{\delta }}^i\left(\hat{A}\right)}_v\left(\lambda \right),{{\breve{\delta }}^{o}\left(\hat{A}\right)}_v\left(\lambda \right)) 
\end{equation}
defines the interval or range of possible values of the quantity $A$ [45]. In quantum mechanics, we deal with an open system including observer, observed system, and measurement and a relative frequency interpretation is used. Therefore, the probability is a fundamental concept. But in topos quantum mechanics, for both open and closed systems, we use a logical interpretation in terms of truth values ($\check{\delta }\left(\hat{A}\right)$) and in consequence, probability is an emergent concept [45]. 

Therefore, in topos language, we have
\[{Network}_{spacetime}\to M_Q\to \underline{\mathrm{\Sigma }}\to {\underline{R}}^{\leftrightarrow }{\ ;\ R}^n\leftarrow M\leftarrow M_Q{\leftarrow Network}_{spacetime}\] 
with the Alexandrov algebra for $\underline{\mathrm{\Sigma }}\to {\underline{R}}^{\leftrightarrow }$ and discrete calculus for $M\to R^n$. Here, the  operators and states are shown by matrices but ${{\delta }^i(\hat{A})}_v$ and ${{\delta }^o(\hat{A})}_v$ are real numbers. Now, $M_Q$ with the above definition is called topos discrete quantum manifold such that both maps $f:$ $\underline{\mathrm{\Sigma }}\to {\underline{R}}^{\leftrightarrow }$ and $g:M\to R^n$ are functions in classical snapshot of a quantum system [45]. Based on this interpretation, time is understood as the difference of measurable values between the classical snapshot of the network with $N$-nodes and the classical snapshot of the network with $(N+1)$-nodes. If there is no change, time has no meaning and cannot be understood [26, 48].

\section{Lorentz Transformation}

\noindent At Planck scale, the space-time is discrete [21-24], and there exist regions of space-time with the smallest area and volume which cannot be divided into smaller regions and are described by spin foams [4, 14, 15].  Alternatively, it can be shown that, the geometry could be decomposed into triangular chunks or their higher-dimensional versions [49]. Therefore, the notion of point is not regarded as fundamental any more but is replaced by the notion of a region which physical objects occupy. Therefore, the notion of a point in space-time becomes secondary to the notion of a region and a description of space-time in terms of extended region might be more appropriate [50]. In topos approach, the notion of a space-time point is replaced by the notion of a space-time region. Such regions should be interpreted as defining regions which are occupied by extended objects. The interesting feature is that the collection of such regions carry a Heyting algebra structure, which is generalized Boolean algebra where the law of excluded middle (for every~proposition,~either~this proposition or its~negation~is~true) does not hold [46]. Therefore, space-time is modeled via an algebra of open regions of space-time where the algebraic operations are interpreted as defining unions and intersections of space-time regions. The reason why open regions are preferred is to account for generalized uncertainty principle [46].

In above, we saw that ${{\underline{R}}^{\leftrightarrow}}_v$ are composed by ${\breve{\delta }\left(\hat{A}\right)}_{v}=({{\breve{\delta }}^i\left(\hat{A}\right)}_v\left(\lambda \right),{{\breve{\delta }}^o\left(\hat{A}\right)}_v\left(\lambda \right))$ where, $\lambda $, ${{\breve{\delta }}^i\left(\hat{A}\right)}_v$, and ${{\breve{\delta }}^o\left(\hat{A}\right)}_v$ are the set of eigenvalues of the associated self-adjoint operator $\hat{A}$, order reversing function, and order preserving function, respectively [46]. Therefore, for each $v_1$ and $v_2\in V(\mathcal{H})$, such that $v_1\subseteq v_2$, and both represent a classical snapshots of the quantum system, an intervals of real numbers ${\breve{\delta }\left(\hat{A}\right)}_{v_1},\ {\breve{\delta }\left(\hat{A}\right)}_{v_2}\in {{\underline{R}}^{\longleftrightarrow }}_{v_1,v_2}$ are referred. If $\hat{A}\in v_2$ and $\hat{A}\notin v_1$, we will have to approximate $\hat{A}$ so as to fit $v_1$. Such an approximation will inevitably coarse-grain $\hat{A}$ i.e., it will deform it [46]. Since, $v_1\subseteq v_2$ then
\begin{equation}
{\breve{\delta }\left(\hat{A}\right)}_{v_1}=\left[c,d\right]\supseteq \left[a,b\right]={\breve{\delta }\left(\hat{A}\right)}_{v_2}(c\le a,\ d\ge b)
\end{equation}
 i.e., the context of $v_1$ are less precise than $v_2$ and in consequence have less information due to coarse-graining [46]. If these intervals are to be interpreted as the regions of space-time which physical objects occupy, by going to a smaller context $v_1$, the precision of determining the position of physical objects decreases [46]. It is possible to view ${{\underline{R}}^{\leftrightarrow}}_v$ as sub-object of a locale in sheaves of $V(\mathcal{H})$ ($Sh(V\left(\mathcal{H}\right))$ [26, 45, 46]. 

Manifolds with metric (relativistic space-time) can be straightforwardly understood as general relativity category (\textbf{GR}) whose morphism are isometric embedding, which are the accepted standard morphism for relativistic space-time [51]. If $A$ be a smooth algebra and $\hat{g}$ be a metric on $A$ of Lorentz signature, a pair $(A,\hat{g})$ is called Einstein algebra [51]. The objects of category of Einstein algebras ($\boldsymbol{EA}$) are $(A,\hat{g})$ and its arrows are Einstein algebra homomorphisms [51]. It can be shown that $\boldsymbol{EA}$\textbf{ }and \textbf{GR} are dual categories [51]. Also, it is well known that if $A$ be a frame of reference and $B$ be a frame of reference moving with velocity $\overrightarrow{v}$ in relation to $A$, a Lorentz transformation is a function ${\mathrm{\Lambda }}_{\overrightarrow{v}}:A\to B$, defined by
\begin{equation}
{\mathrm{\Lambda }}_{\overrightarrow{v}}\left(t,\overrightarrow{x}\right)={\gamma }_{\overrightarrow{v}}(t-\frac{\left\langle \overrightarrow{v},\overrightarrow{x}\right\rangle }{c^2},\overrightarrow{x}-\overrightarrow{v}t) 
\end{equation}
where $\left\langle \overrightarrow{v},\overrightarrow{x}\right\rangle $ represents the scalar product in $R^3$. The transformation has linearity property, ${\mathrm{\Lambda }}_{\overrightarrow{0}}=id_A$, ${\mathrm{\Lambda }}_{\overrightarrow{u}}\circ {\mathrm{\Lambda }}_{\overrightarrow{v}}={\mathrm{\Lambda }}_{\overrightarrow{u}+\overrightarrow{v}}$, and ${\mathrm{\Lambda }}_{-\overrightarrow{v}}={({\mathrm{\Lambda }}_{\overrightarrow{v}})}^{-1}$ [52]. Then, the Lorentz category is defined as the category whose objects are inertial frames of reference and morphism are Lorentz transformations [52]. 

Similarly, we can define the Lorentz category, $\mathcal{L}$, as follows:

\noindent (i): Objects: ${{\underline{R}}^{\leftrightarrow }}_{V}$ $=\{(\mu ,\nu )~\vert\mu ,\ \nu :\downarrow V\to $ $R~\vert\ ,$

\noindent $\ \ \ \ \ \ \ \ \ \ \ \ \ \ \ \ \ \ \ \ \ \ \ \ \ \ \ \ \ \ \ \ \ \ \ \ \ \ \mu $ is order preserving, $\nu $ is order reversing; $\mu \le \nu \}$ 

\noindent which ${{\underline{R}}^{\leftrightarrow }}_V$ is a locale space-time and is seen as a collection of unions of varying intervals of real numbers and $V\in V(\mathcal{H})$ represents a classical snapshot of the quantum system.

\noindent (ii): Morphisms: ${\mathrm{\Lambda }}_{V_1,V_2}:\ {{\underline{R}}^{\leftrightarrow }}_{V_1}\to {{\underline{R}}^{\leftrightarrow }}_{V_2}$ , $\left({\mu }_1,v_1\right)\longmapsto {(\mu }_2,v_2)$ which $\mathrm{\ }{\mathrm{\Gamma }}_{V_1,V_2}$ is the Lorentz transformation i.e., ${\mathrm{\Lambda }}_{V_1,V_2}=$$\mathrm{\{}$${\mu }_1\longmapsto {\mu }_2$ and $v_1\longmapsto v_2$ $\mathrm{\}}$. 

\noindent (iii) Let us, assume $({\mu }_1,v_1)\in $ ${{\underline{R}}^{\leftrightarrow }}_{V_1}\ $moves with velocity $\overrightarrow{v}$ in relation to $\left(\mu ,v\right)\in \ {{\underline{R}}^{\leftrightarrow }}_V$ and $({\mu }_2,v_2)\in \ {{\underline{R}}^{\leftrightarrow }}_{V_2}$ moves with velocity $\overrightarrow{u}$ in relation to $({\mu }_1,v_1)\in $ ${{\underline{R}}^{\leftrightarrow }}_{V_1}$. If ${\mathrm{\Lambda }}_{\overrightarrow{u}}\circ {\mathrm{\Lambda }}_{\overrightarrow{v}}={\mathrm{\Lambda }}_{\overrightarrow{u}+\overrightarrow{v}}$ satisfies, then there will be exist a morphism ${\mathrm{\Lambda }}_{\overrightarrow{u}+\overrightarrow{v}}:{{\underline{R}}^{\leftrightarrow }}_V\to {{\underline{R}}^{\leftrightarrow }}_{V_2}$ , $\left(\mu ,v\right)\longmapsto {(\mu }_2,v_2)$.

\noindent (iv) If, in case (iii), $({\mu }_3,v_3)\in $ ${{\underline{R}}^{\leftrightarrow }}_{V_3}$ moves with velocity $\overrightarrow{w}$ in relation to $({\mu }_2,v_2)\in $ ${{\underline{R}}^{\leftrightarrow }}_{V_2}$. Then ${\mathrm{\Lambda }}_{\overrightarrow{u}+\overrightarrow{v},\overrightarrow{w}}:{{\underline{R}}^{\leftrightarrow }}_{V_2}\to {{\underline{R}}^{\leftrightarrow }}_{V_3}$ , $({\mu }_2,v_2)\longmapsto {(\mu }_3,v_3)$. However, ${\mathrm{\Lambda }}_{\overrightarrow{u}}:{{\underline{R}}^{\leftrightarrow }}_{V_1}\to {{\underline{R}}^{\leftrightarrow }}_{V_2}$ , $\left({\mu }_1,v_1\right)\longmapsto {(\mu }_2,v_2)$ and ${\mathrm{\Lambda }}_{\overrightarrow{w}}:{{\underline{R}}^{\leftrightarrow }}_{V_2}\to {{\underline{R}}^{\leftrightarrow }}_{V_3}$ , $\left({(\mu }_2,v_2\right)\longmapsto ({\mu }_3,v_3)$ , then ${\mathrm{\Lambda }}_{\overrightarrow{u}+\overrightarrow{w}}:{{\underline{R}}^{\leftrightarrow }}_{V_1}\to {{\underline{R}}^{\leftrightarrow }}_{V_3}$ , $\left({\mu }_1,v_1\right)\longmapsto {(\mu }_3,v_3)$. Therefore, ${\mathrm{\Lambda }}_{\overrightarrow{u}+\overrightarrow{w},\overrightarrow{v}}:{{\underline{R}}^{\leftrightarrow }}_V\to {{\underline{R}}^{\leftrightarrow }}_{V_3}$ , $(\mu ,v)\longmapsto {(\mu }_3,v_3)$.

\noindent (v) if, ${\mathrm{\Lambda }}_{\overrightarrow{0}}=id_A$, then ${\mathrm{\Lambda }}_{\overrightarrow{u}+\overrightarrow{0}}:{{\underline{R}}^{\leftrightarrow }}_V\to {{\underline{R}}^{\leftrightarrow }}_V$ , $\left(\mu ,v\right)\longmapsto (\mu ,v)$. 

But, a question can be asked: how ${\mathrm{\Lambda }}_{V_1,V_2}=$$\mathrm{\{}$${\mu }_1\longmapsto {\mu }_2$ and $v_1\longmapsto v_2$ $\mathrm{\}}$ acts? Or, how the Lorentz transformation acts on the locale space-time? For given transformation group $G$ and for each $V\in V(\mathcal{H})$, one can consider the collection of all homomorphisms ${\phi }_g:\downarrow V\to V\left(\mathcal{H}\right),\ g\in G$, such that ${\phi }_g\left(V\right)={\hat{U}}_gV{\hat{U}}_{g^{-1}}$, for some unitary representation
 ${\hat{U}}_g$ of $g$ [45]. Therefore, ${{\underline{R}}^{\longleftrightarrow }}_V$ which are composed by ${\breve{\delta }\left(\hat{A}\right)}_V=({{\check{\delta }}^i\left(\hat{A}\right)}_V\left(\lambda \right)=\mu ,{{\check{\delta }}^o\left(\hat{A}\right)}_V\left(\lambda \right)=v)$ and live in $V\in \ \downarrow V$ are transferred under acting a unitary transformation as
\begin{equation}
\check{\delta }{({\hat{U}}_g\hat{A}{\hat{U}}^{-1}_g)}_{l_g(V)}=({\delta }^i{\left({\hat{U}}_g\hat{A}{\hat{U}}^{-1}_g\right)}_{l_g\left(V\right)}={\mu }_1,\ {\delta }^o{\left({\hat{U}}_g\hat{A}{\hat{U}}^{-1}_g\right)}_{l_g\left(V\right)}=v_1)
\end{equation}
 which living in the transformed new context $l_g(V)$. Since, generally $\hat{A}\neq {\hat{U}}_g\hat{A}{\check{U}}^{-1}_g$ and $V\neq l_g(V)$, the locale space-time is not invariant under given transformation group $G$ e.g. Lorentz transformation.

\section{Cosmological Constant}

\noindent It is well known that, the Einstein equations can be found by using the action [53]
\begin{equation}
S=\frac{1}{16\pi G}\int{\left(R-2\mathrm{\Lambda }\right)\sqrt{-g}d^4x+\int{L_{matter}(\phi ,\partial \phi )\sqrt{-g}d^4x}} 
\end{equation}
The gravitational field is described by the Lagrangian, $L_{grav}\propto (\frac{1}{G})(R-2\mathrm{\Lambda })$, which interacts with matter described by the Lagrangian $L_{matter}(\phi ,\partial \phi )$ [53]. Therefore, gravity is described by the Newton constant $G$ and the cosmological constant $\mathrm{\Lambda }$ [53]. Using $L_{grav}$, one can describes $\mathrm{\Lambda }$ as the action per unit space-time volume which is due just to the existence of space-time as such, independent of matter or gravitational waves i.e.,
\begin{equation}
S_{\mathrm{\Lambda }}=\frac{-\mathrm{\Lambda }}{8\pi G}\int{\sqrt{-g}d^4x=\frac{-1}{8\pi G}\int{\mathrm{\Lambda }\mathrm{\ }d\mathcal{V}}=\frac{-\mathrm{\Lambda }}{8\pi G}\mathcal{V}} 
\end{equation}
and therefore $\mathrm{\Lambda }=\frac{-1}{\mathcal{V}}S_{\mathrm{\Lambda }}$ ($8\pi G=1$) i.e., $\mathrm{\Lambda }$ and $\mathcal{V}$ are canonical conjugates [53]. Also, if the space-time is homogeneous, isotropic, and spatially flat i.e.,
\begin{equation}
ds^2=-dt^2+a{\left(t\right)}^2(dx^2+dy^2+dz^2)
\end{equation}
the Einstein equations reduce to a pair of ordinary differential equations for the scale factor $a$. The first equation, which is called Friedmann equation or Hamiltonian constraint, is [54, 55]
\begin{equation}
{(\frac{\dot{a}}{a})}^2=\frac{1}{3}\rho +\frac{\mathrm{\Lambda }}{3}
\end{equation}
and the second equation, which involves $\ddot{a}$ , is
\begin{equation}
\frac{\ddot{a}}{a}=-\frac{1}{6}\left(\rho +3p\right)+\frac{\mathrm{\Lambda }}{3}
\end{equation}
It has been shown that the second equation is unstable under small fluctuations from the true solution and in consequence cannot be used as dynamical guide for everpresent $\mathrm{\Lambda }$ model [54]. Also, the combination of these two equations is unstable and cannot be used as dynamic guide [54, 55]. 

In causal set theory, the fluctuations of cosmological constant, $\mathrm{\Lambda }$, arise from the underlying space-time discreteness [54, 55]. There are four inputs to the argument which are: (i) the finite number $N$ of elements is due to the space-time discreteness, (ii) the space-time volume, $\mathcal{V}$, is determined by the number of elements, (iii) $\mathrm{\Lambda }$ and $\mathcal{V}$ are conjugated, and (iv) the fluctuation in $\mathcal{V}$ is due to the Poisson fluctuations in $N$ [54]. However, in order to do justice to local Lorentz invariance, the corresponding between $N$ and $\mathcal{V}$ cannot be exact, but should be subject to Poisson fluctuations with a typical magnitude of $\sqrt{N}$. It should be noted that, for Poisson distribution $f\left(\lambda ,N\right)=\frac{e^{-N}N^{\lambda }}{\lambda !}$, the mean value and standard deviation are $N$ and $\sqrt{N}$, respectively. Therefore, for fixing $N$ at the fundamental level, one can fix $\mathcal{V}$ only up to fluctuations of magnitude $\pm \sqrt{\mathcal{V}}$ , i.e., $N\sim \mathcal{V}\pm \sqrt{\mathcal{V}}\ $[54, 55]. Using Heisenberg uncertainty principle, and since $\mathrm{\Lambda }$ and $\mathcal{V}$ are conjugated, one obtains $\Delta \mathrm{\Lambda }\times \Delta \mathcal{V}\sim 1$ ($\hslash =1$) and in consequence $\mathrm{\Lambda }\sim \Delta \mathrm{\Lambda }\sim {\frac{1}{\Delta \mathcal{V}}\sim \pm \mathcal{V}}^{-1/2}$ in natural units [54, 55]. Assuming $\left\langle \mathrm{\Lambda }\right\rangle =0$ and taking $\mathcal{V}$  to be roughly the fourth power of the Hubble radius, $H^{-1}$, then $\mathrm{\Lambda }\sim H^2=\frac{1}{3}{\rho }_c$ ($8\pi G=1$), where ${\rho }_c$ is the critical density of the universe [54, 55]. Therefore, causal set theory predicts that $\mathrm{\Lambda }$ is of order $H^2$ and is everpresent and also fluctuates due to Poisson statistic of space-time causal elements [54, 55]. In consequence, $\mathrm{\Lambda }\sim \Delta \mathrm{\Lambda }\sim {\frac{1}{\Delta \mathcal{V}}\sim \pm \mathcal{V}}^{-1/2}\sim \pm {10}^{-12}$ in natural units [54, 55, 56]. Since, $\mathrm{\Lambda }\mathrm{(}=\frac{-1}{V}S_{\mathrm{\Lambda }}$) fluctuates due to Poisson statistic of space-time causal elements and the space-time volume can be re-interpreted as the number of elements, then $\mathrm{\Lambda }$ is the action per element and one can write [55]
\begin{equation}
\Delta S_{\mathrm{\Lambda }}=\beta \xi \sqrt{N\left(t+1\right)-N(t)}
\end{equation}
It has been shown that $0.01<\beta <0.02$ and $\xi $ is a random number which drives the random walk [55]. 

As we have mentioned before, in topos approach, the notion of a space-time point is replaced by the notion of a space-time region, called locale, i.e., ${{\underline{R}}^{\leftrightarrow }}_V$ is a locale space-time and is seen as a collection of unions of varying intervals of real numbers and $V\in V(\mathcal{H})$ represents a classical snapshot of the quantum system. It is remembered that, if the state $\psi $ is an eigenstate of the physical quantity $A$, then one would get a sharp value of the quantity $A$, say $a$. If $\psi $ is not an eigenstate, then he/she would get a certain range $\mathrm{\Delta }$ of values for $A$, where $\mathrm{\Delta }\in sp(\hat{A})$, and $sp$ stands for spectrum. In topos language, it means that $\hat{A}$ gets approximated  values both from above, through the process of outer daseinisation (${\check{\delta }}^o{(\hat{A})}_V$) and from below, through the process of inner daseinisation (${\check{\delta }}^i{(\hat{A})}_V$) [26, 45].  Such an approximation gets coarser as $W\subseteq V$ ( gets smaller), which basically means that $W$ contains less and less projections, i.e., less and less information [26, 45]. In this way, the point particle problem is also solved and probability is an emergent concept. Therefore, if the space-time network includes $N$-event (point) we can attribute to each point a locale ${{\underline{R}}^{\leftrightarrow }}_i,i=1,2,\cdots ,N$, at each classical snapshot, $j$. In consequence, we can write
\begin{equation}
 ({Volume)}_j\sim \sum^{N_1}_{i\in j}{{{\underline{R}}^{\leftrightarrow }}_i}
\end{equation}
and
\begin{equation}
\Delta {(Volume)}_{j,j+1}\sim \{\sum^{N_2}_{i\in j+1}{{{\underline{R}}^{\leftrightarrow }}_i}-\sum^{N_1}_{i\in j}{{{\underline{R}}^{\leftrightarrow }}_i}\} 
\end{equation}
The fluctuation in volume $\mathcal{V}$ is due to the fluctuations in both $N$ and ${{\underline{R}}^{\leftrightarrow }}_i$ (due to the variations in (${\check{\delta }}^i{(\hat{A})}_V,{\check{\delta }}^o{(\hat{A})}_V$) and algebra $V$). Since, $\mathrm{\Lambda }$ and $\mathcal{V}$ are canonical conjugates, one can write $\mathrm{\Lambda }\sim \Delta \mathrm{\Lambda }\sim \frac{1}{\Delta \mathcal{V}}$. Therefore, $\mathrm{\Lambda }$ is not zero and fluctuates ($\Delta \mathrm{\Lambda }$) due to the fluctuations in the in both $N$ and ${{\underline{R}}^{\leftrightarrow }}_i$. If the fluctuation in volume $\mathcal{V}$ obeys Poisson fluctuations (it is assumed, without proof), then $\mathrm{\Lambda }\sim \Delta \mathrm{\Lambda }\sim {\frac{1}{\Delta \mathcal{V}}\sim \pm \mathcal{V}}^{-1/2}$. Similar to the causal set theory, assuming $\left\langle \mathrm{\Lambda }\right\rangle =0$ and taking volume to be roughly the fourth power of the Hubble radius, $H^{-1}$, then $\mathrm{\Lambda }\sim H^2=\frac{1}{3}{\rho }_c$. Therefore, similar to the causal set theory, this topos-based theory predicts that the cosmological constant, $\mathrm{\Lambda }$ , fluctuates due to the fluctuation in volume $\mathcal{V}$ due to the both $N$ and ${{\underline{R}}^{\leftrightarrow }}_i$and can be of order $H^2$, and everpresent.

\section{Inflation of Universe}

In causal set theory, a new causal set can be formed by adjoining a single maximal element to a given causet. Each given causet and each new causet are called parent and child, respectively. If the adjoined element is timelike to every element of the parent it will be called the timid child and if it is spacelike to every element of the parent it will be called gregarious child. Each parent-child relationship describes a transition, from one causal set ($C_1$) to the another ($C_2$) induced by birth of a new element ($C_1\to C_2$). $C_1$ is referred as the precursor of the transition.  If $C_n$ stands for the set of causets with $n$ elements, the set of all transitions from $C_n$ to $C_{n+1}$ is called stage $n$ [57, 58]. In General, a classical sequential growth (CSG) model is specified by a countable set of non-negative constants $t_0,\ t_1,\ t_2,\ t_3,\ t_4,\ \cdots $, where $t_k$ is the relative probability that the newly born element choose a particular set of ancestors which has cardinality $k$. The probability of the transition $C_n\to C_{n+1\ }$is $\frac{\lambda (\overline{\omega },m)}{\lambda (n,0)}$, where $\lambda \left(\overline{\omega },m\right)=\sum^{\overline{\omega }}_{k=m}{( \begin{array}{c}
\overline{\omega }-m \\ 
k-m \end{array}
)t_k}$, $\overline{\omega }$ is the cardinality of the precursor and $m$ is the number of maximal elements of the precursor [59]. It has been shown that, in a CSG model ($t_{k_1}$), if the causal set that grows contains a poset (a Big Crunch--Big Bang event) the effective dynamics of the causal set after the poset is governed by a different CSG model with a renormalized set of constants $\mathrm{\{}$$t_{k_2}$$\mathrm{\}}$ [59, 60].

 If the transition probability ${\alpha }_n=\frac{\lambda (\overline{\omega },m)}{\lambda (n,0)}\ $from $C_n$ to a specified causet $C_{n+1}$ of size $n+1$ is given by ${\alpha }_n=p^m{(1-p)}^{n-\overline{\omega }}$, where $m$ is the number of maximal elements in the precursor set and $\overline{\omega }$ is the size of the entire precursor set, the randomly growing causet is called transitive percolation [57, 58, 61]. It is obvious that the gregarious transition will occur with the probability $q_n=q^n$, where $q=1-p$. Finally, it can be shown that ${\alpha }_n=\sum^m_{k=0}{{\left(-\right)}^k( \begin{array}{c}
m \\ 
k \end{array}
)\frac{q_n}{q_{\overline{\omega }}-k}}$ [57, 58]. It is a very general family of classically stochastic, sequential growth dynamics for causal sets and illustrate how non-gravitational matter can arise dynamically from the causal set without having to be built in at the fundamental level [57, 58, 61]. Another more interesting case is when $t_k=t^k/k!$. It can be shown that ${\lambda }_{eff}={(\frac{t}{\sqrt{N}})}^m\mathrm{exp}\mathrm{}(\left(\frac{2\overline{\omega }-m}{2}\right)\sqrt{\frac{t}{N}})$ where $N$ is the number of elements in the past of a single element [61]. This model includes two effective couplings $t$ and $1/\sqrt{N}$ and shows how the bouncing could affect the dynamics [61]. 

We have shown [25] that how one can define a new growth mechanism and define an entropy function by mixing the idea about the generations in the causal set theory [58] and the idea about the aged-structure populations model [62]. In the new model, the number of parents increase by increasing the growth steps i.e., the new causal children (timid children) in $i$-th step are considered as new parents of the children in $(i+1)$-th step. Also, it is assumed that the non-causal children (gregarious children) in $i$-th step do not take part in the growth process and survive to $(i+1)$-th step. Therefore, the change in age structure between step $i$ and step $(i+1)$ can be found by using the Leslie matrix ($M$) [25]. If generally, $m_i$ stands for the number of causal children in the $i$-th step respect to step one, $b_i$ stands for the proportion of non-causal children in $i$-th step surviving to $(i+1)$-step, and $l_j=\prod^{j-1}_{r=1}{b_r}$ for $j\ge 2$, $l_1=1$, one can solve two eigenvalue equations $M\overrightarrow{u}=\lambda \overrightarrow{u}$ and $\overrightarrow{v}M=\lambda \overrightarrow{v}$ and finds $u_i=l_i/{\lambda }_i$ and $v_i=(\sum^n_{j=i}{m_ju_j)/u_i}$ [25]. By defining the probability element ($p_i$) and the probability matrix (P$=(P_{ij})$) as $p_i=m_il_i/{\lambda }_i$ and $P_{ij}=\{p_i\ \left(i=1\right);\ 1\ \left(i=j=1\right);0\ \left(otherwise\right)\}$, respectively, one can define the population entropy for the stationary distribution as $S=-\frac{\sum^n_{i=1}{p_i\mathrm{log}\mathrm{}(p_i)}}{\sum^n_{i=1}{ip_i}}$ [25]. It can be shown that $\mathrm{\Delta }S={(\sum^n_{i=1}{ip_i)}}^{-1}\sum_i{V_i>0}$  where the amount of $\Delta \mathrm{\Phi }\ (\mathrm{\Phi }=\frac{\sum^n_{i=1}{p_ilog{\lambda }_ip_i}}{\sum^n_{i=1}{ip_i}})$ is increased by $V_i$. In consequence, the entropy increases by growing the population and in consequence the population will be more stable than before. It should be noted that $\Delta \mathrm{\Phi }$ is similar to the internal mean energy [25]. If it is assumed that ${\sum_i{V}}_i=m\hslash \omega /2$, where $m$ is the number of causal parents which took part in the evolution of the system, then the entropy is quantized and equal to $S=+\frac{m\hslash }{2}{(\sum^n_{i=1}{ip_i)}}^{-1}$ where $T={(\sum^n_{i=1}{ip_i)}}^{-1}$ is constant [25]. Then, why we see universe inflation because its entropy increases due to the increment of the space-time events [25]. 

Now, we assume that the growth of the causal set is done only through the birth of timed children, since there are placed on the worldline and there is the causality relation between them. Also let us, consider an arbitrary causal set $C_n$ of size $n$. It is assumed that in the causet, there are $m$ number of elements to which a new element can join and a new causet $C_{n+1}$ is created in this way. Now, we encounter the proposition ($A\in \Delta $): the transition probability $\left(C_n\to C_{n+1}\right)\in r=(0,1]$ i.e., in particular, what it represents, is the truth value for the proposition ($A\in \Delta $) to be true with the probability at least $r$ [45]. The value $r=0$ is not included to avoid obtaining situations in which all propositions are totally true (even, the totally false propositions) with probability zero [45].  Now, a question can be asked: how can one relate the truth value to probability measure? In order to follow this discussion better, let us review a simple example. If there are three precursor elements  $\mathrm{\Omega }\mathrm{=}$ $\mathrm{\{}$$1,\ 2,\ 3\}$ i.e., $m=3$, and each with $p=1/3$, then the probability (state) space includes $2^3$ elements i.e.,
\begin{equation}
\mathrm{\Sigma }=\{\{\};\{1\};\{2\};\{3\};\{1,2\};\{1,3\};\{2,3\};\{1,2,3\}\}
\end{equation}
The state space can be mapped to unit interval $[0,1]$ through a probability measure map $\mu :Sub(\mathrm{\Sigma }\mathrm{)}\to [0,1]$, where $Sub(\mathrm{\Sigma }\mathrm{)}$ denotes the measurable subset of$\mathrm{\ }\mathrm{\Sigma }$. Therefore,
\begin{equation}
\{\}\to 0, \{1\}\to 1/3, \{2\}\to 1/3, \{3\}\to 1/3, \{1,2\}\to 2/3,  
\end{equation}
\begin{equation}
\{1,3\}\to 2/3, \{2,3\}\to 2/3, \{1,2,3\}\to 1 
\end{equation}

It is now possible to define a classical measure dependent truth object as: ${\mathbb{T}}^{\mu }_{r}=\{S\subseteq \mathrm{\Sigma }\vert\mu (S)\ge r\}$ for all $r\in (0,1]$ [45]. What the above truth object defines are all those propositions which are true with probability equal or greater than $r$ [45]. However, we would like to find their analogues in topos $Sh(({0,1)}_L)$ where $Sh$ stands for sheaf and $({0,1)}_L$ is the topological space.

For doing that, we map the state space $\mathrm{\Sigma }$ to the constant sheaf $\underline{\mathrm{\Sigma }}$ i.e., $\mathrm{\Sigma }\to \underline{\mathrm{\Sigma }}$, which means that for all $(0,r)\in \mathcal{O}(({0,1)}_L)$ then $\underline{\mathrm{\Sigma }}\left(0,r\right)=\mathrm{\Sigma }$, where $\mathcal{O}(({0,1)}_L)$ is the collection of open sets. Also, for any measurable set $S\in Sub(\mathrm{\Sigma })$ the constant sheaf $S\to \underline{S};\ \underline{S}\left(0,r\right)=S$ can be defined for obtaining the map [45]
\begin{equation}
\Delta :Sub\left(\mathrm{\Sigma }\right)\to {Sub}_{Sh(({0,1)}_L)}\left(\underline{\mathrm{\Sigma }}\right);S\to \underline{S}
\end{equation} 

Then, the analogue of the truth object in $Sh(({0,1)}_L)$ is [45]
\begin{equation}
{\underline{\mathbb{T}}}^{\mu }_{(0,r)}=\{\underline{S}\subseteq \underline{\mathrm{\Sigma }}\vert\mu (\underline{S})\ge r\} for all (0,r)\in \mathcal{O}(({0,1)}_L)
\end{equation}
What is about the quantum probability?

If $\mathcal{V}(\mathcal{H})$ be the poset of all abelian von Neumann algebras of a given Hilbert space $\mathcal{H}$, the presheaf
\begin{equation} 
{\underline{[0,1]}}^{\succcurlyeq }:\mathcal{V}(\mathcal{H})\to Sets
i.e, {\underline{[0,1]}}^{\succcurlyeq }_{V}=\{V\in \mathcal{V}\left(\mathcal{H}\right);\ f:\downarrow V\to [0,1]\} 
\end{equation}
which $f$ is order reversing, can be defined [45]. Here, $\downarrow V$ means the lower set of  $V$. A measure $\mu $ on the state space $\underline{\mathrm{\Sigma }}$  is a map
\begin{equation}
\mu :{Sub}_{cl}(\underline{\mathrm{\Sigma }})\to \mathrm{\Gamma }{[0,1]}^{\succcurlyeq }; \underline{S}={({\underline{S}}_V)}_{V\in \mathcal{V}(\mathcal{H})}\mapsto {(\mu \left({\underline{S}}_V\right))}_{V\in \mathcal{V}(\mathcal{H})}
\end{equation}
where ${Sub}_{cl}(\underline{\mathrm{\Sigma }})$ stands for both closed and open sub-objects of $\underline{\mathrm{\Sigma }}$ and $\mathrm{\Gamma }\left(\mathrm{\Omega }\right)$ is the collection of all global elements of $\mathrm{\Omega }$ which forms a Heyting algebra [45]. It is remembered that, a terminal object in category $C$ is a $C$-object $1$ such that, given any other $C$-object $A$, there exist one and only one $C$-arrow from $A$ to $1$ ($l:A\to 1$). In a post the terminal element object is the greatest element with respect to the ordering. A global section or global element of a presheaf $X$ is an arrow, $k:1\to X$, from the terminal object $1$ to the presheaf $X$ [45]. It has been shown that, for the defined measure $\mu $, there exist a unique state $\rho $ associated to that measure [45]. What is about truth object and truth value?

A truth object, for a given context $V$, is the sheaf
\begin{equation}
{\underline{\mathbb{T}}}^{\left.\vert\psi \right\rangle }_V:=\{\underline{S}\in {Sub}_{cl}\left({\underline{\mathrm{\Sigma }}}_{\vert\downarrow V}\right)\vert for\ V_1\subseteq V\in \mathcal{V}\left(\mathcal{H}\right),\}
\end{equation}
\begin{equation}
\left\langle \psi \mathrel{\left|\vphantom{\psi  {\hat{P}}_{S_{V_1}} \psi }\right.\kern-\nulldelimiterspace}{\hat{P}}_{S_{V_1}}\mathrel{\left|\vphantom{\psi  {\hat{P}}_{S_{V_1}} \psi }\right.\kern-\nulldelimiterspace}\psi \right\rangle =1\ i.e.,\ \left.\vert\psi \right\rangle \left\langle \psi \vert\right.\le {\hat{P}}_{S_{V_1}}
\end{equation}
Here, $\mathrm{\vert}$$\left.\psi \right\rangle $ and ${\hat{P}}_{S_{V_1}}\in P(V_1)$ stand for state and projection operator, respectively [45].  The truth value now becomes
\begin{equation}
{v\left(A\in \Delta ,\ \left.\vert\psi \right\rangle \right)}_V={[\left[\ \underline{\delta \left(\hat{E}\left[A\in \Delta \right]\right)}\in {\underline{\mathbb{T}}}^{\vert\left.\psi \right\rangle }\right]]}_V 
\end{equation}
where $\hat{P}=\hat{E}\left[A\in \Delta \right]=\vert\left.\psi \right\rangle \left\langle \psi \vert\right.$ is the corresponding projection operator in terms of the true object which represents the proposition $A\in \Delta $ [45]. 
If we deal with the density matrix $\rho $ then [45]
\begin{equation}
{\underline{\mathbb{T}}}^{\rho ,r}_V:=\{\underline{S}\in {Sub}_{cl}\left({\underline{\mathrm{\Sigma }}}_{\vert\downarrow V}\right)\vert for\ V_1\subseteq V\in \mathcal{V}\left(\mathcal{H}\right),\ {tr(\rho \hat{P}}_{S_{V_1}})\ge r\}
\end{equation}
  and
\begin{equation}   
v{\left(A\in \Delta ,\rho \right)}^r\left(V\right)={\left[\left[\ \underline{\delta \left(\hat{E}\left[A\in \Delta \right]\right)}\in {\underline{\mathbb{T}}}^{\rho ,r}\right]\right]}_V=
\end{equation}
\begin{equation}
\{V_1\subseteq V\ tr{(\rho \hat{E}\left[A\in \Delta \right])}_{V_1}\ge r\}
\end{equation}
Finally, it can be shown that probabilities can be replaced by truth values without any information being lost i.e, we can effectively replace the probability measure with the collection of truth values $\mathrm{\Gamma }{\underline{\mathrm{\Omega }}}^{Sh\left({\left(0,1\right)}_L\right)}$ for classical probability and $\mathrm{\Gamma }({\underline{\mathrm{\Omega }}}^{sh{(\mathcal{V}(\mathcal{K})\times (0,1)}_L)})$ for quantum probability [45]. It means that there is the following commutative diagram for classical probability
\[Sub\left(X\right)\to [0,1]\to \mathrm{\Gamma }{\underline{\mathrm{\Omega }}}^{Sh\left({\left(0,1\right)}_L\right)}\leftarrow {Sub}_{Sh\left({\left(0,1\right)}_L\right)}(\underline{X})\leftarrow Sub(X)\] 
and the following commutative diagram for quantum probability
\[Sub\left(X\right)\to [0,1]\to \mathrm{\Gamma }{\underline{\mathrm{\Omega }}}^{Sh\left(\mathcal{V}(\mathcal{H})\times {\left(0,1\right)}_L\right)}\leftarrow {Sub}_{Sh\left({\left(0,1\right)}_L\right)}(\underline{X})\leftarrow Sub(X)\] 
where $\mu :Sub\left(X\right)\to [0,1]$ stands for probability measure
\begin{equation}
{\epsilon }^{\mu }:{Sub}_{Sh\left({\left(0,1\right)}_L\right)}(\underline{X})\to \mathrm{\ }\mathrm{\Gamma }{\underline{\mathrm{\Omega }}}^{Sh\left({\left(0,1\right)}_L\right)}
\end{equation}
and 
\begin{equation}
{\epsilon }^{\mu }:{Sub}_{Sh\left({\left(0,1\right)}_L\right)}(\underline{X})\to \mathrm{\Gamma }{\underline{\mathrm{\Omega }}}^{Sh\left(\mathcal{V}(\mathcal{H})\times {\left(0,1\right)}_L\right)} 
\end{equation}
stand for classical and quantum probabilities, respectively [45]. Therefore, there is a bijective correspondence between $\mu $ and ${\epsilon }^{\mu }$ and in consequence,  the probability measure is effectively replaced with the collection of truth values [45]. 

 But, it is known that, each context $V\in \mathcal{V}(\mathcal{H})$ can be generated from a set of pairwise orthogonal projections $\{P_1,P_2,\cdots ,P_k\}$. Therefore, based on the above descriptions, $({\mu }_{\vert V}\left(P_1\right),\ {\mu }_{\vert V}\left(P_2\right),\ \cdots ,\ {\mu }_{\vert V}\left(P_k\right))$ is the probability distribution. Hence to each context $V$ we can assign the Shannon entropy of its associated probability distribution [63]
\begin{equation}
E{(\mu )}_{\vert V}=Sh\left({\mu }_{\vert V}\left(P_1\right),\ {\mu }_{\vert V}\left(P_2\right),\ \cdots ,\ {\mu }_{\vert V}\left(P_k\right)\right)=
\end{equation}
\begin{equation}
-\sum^k_{i=1}{{\mu }_{\vert V}\left(P_i\right)\mathrm{ln}\mathrm{}({\mu }_{\vert V}\left(P_i\right))}
\end{equation}

It should be noted that if the $V$ is a $k$-dimensional context then $E(\mu )\le \mathrm{ln}\mathrm{}(k)$ at $V$, and hence for an $n$-dimensional matrix algebra, $E(\mu )\le \mathrm{ln}\mathrm{}(n)$ at any context [63]. Therefore, the contextual entropy can be seen as a mapping defined on the set of measure associated to a spectral presheaf: $E:\ \mathcal{M}(\underline{\mathrm{\Sigma }})\to \mathrm{\Gamma }{[0,{\mathrm{ln} \left(n\right)\ }]}^{\preccurlyeq }$ [63]. Ofcourse, the definition does not make any direct reference to the quantum state which the measure corresponds to [63]. If $\{P_1,P_2,\cdots ,P_k\}$ is the set of one-dimensional pairwise orthogonal projections, which in turn determine a maximal context $V_{\rho }$ via the double commutant construction, the density matrix $\rho $ is diagonal and ${\lambda }_i=Tr(\rho P_i)$ where ${\{{\lambda }_i\}}^n_{i=1}$ are eigenvalues of $\rho $. Then,
\begin{equation}
E{({\mu }_{\rho })}_{V_{\rho }}=-\sum^n_{i=1}{{\mu }_{\rho }{\vert}_{V_{\rho }}\left(P_i\right){\mathrm{ln\ }{\mu }_{\rho }{\vert}_{V_{\rho }} \left(P_i\right)\ }}=
\end{equation}
\begin{equation}
-\sum^n_{i=1}{Tr\left(\rho P_i\right){\mathrm{ln} Tr\left(\rho P_i\right)=-\sum^n_{i=1}{{\lambda }_i{\mathrm{ln} {\lambda }_i=VN(\rho )\ }}\ }}
\end{equation}
 which is the von Neumann entropy [63]. Therefore, the von Neumann entropy of $\rho $ is equal to the minimal value of the contextual entropy $E_{\rho }(V)$, when $V$ is varying over the set of maximal context [63]. 

As we have mentioned before, in the above classical sequential growth (CSG) model, the probability of the transition $C_n\to C_{n+1\ }$is
\begin{equation}
{\alpha }_n=\prod^m_{i=1}{p_i\prod^{n-\overline{\omega }}_{k=1}{\left(1-p_k\right)=p^m{(1-p)}^{n-\overline{\omega }}}}
\end{equation}
where, $p_i=p$ for all $i$, $p_k=p$ for all $k$, $m$ is the number of maximal elements in the precursor set and $\overline{\omega }$ is the size of the entire precursor set. The set of maximal elements in the precursor set and the set of the entire precursor are shown by $\{e_1$,$e_2,\cdots ,e_m\}\ $ and $\{d_1$,$d_2,\cdots ,d_{n-\overline{\omega }}\}$, respectively. Therefore, $X_1=\{e_1$,$e_2,\cdots ,e_m\}$ is the first state space with the joint probability set $\left\{p_1,p_2,\cdots ,p_m\right\}=\left\{p,p,\cdots ,p\right\}\in [0,1]$ and $X_2=\{d_1$,$d_2,\cdots ,d_{n-\overline{\omega }}\}$ is the second state space with the joint probability set $\left\{1-p_1,1-p_2,\cdots ,1-p_{n-\overline{\omega }}\right\}=\left\{1-p,1-p,\cdots ,1-p\right\}\in [0,1]$. The total state space is $X=X_1+X_2$ with the joint probability set $\{P\}=\left\{p_1,p_2,\cdots ,p_m,1-p_1,1-p_2,\cdots ,1-p_{n-\overline{\omega }}\right\}$. The probability measure is defined as $\mu :X\to \{P\}\in [0,1]$. Now, we map $X$ to the constant sheaf $\underline{X}$  (i.e., $X\to \underline{X}$) and write ${\epsilon }^{\mu }:Sh({\left\{0,1\right\}}_L)(\underline{X})\to \mathrm{\Gamma }{\underline{\mathrm{\Omega }}}^{Sh({\left\{0,1\right\}}_L)}$, where ${\left\{P\right\}}_L$ is the topological space, $Sh({\left\{0,1\right\}}_L)(\underline{X})$ is constant sheaf and $\mathrm{\Gamma }{\underline{\mathrm{\Omega }}}^{Sh({\left\{0,1\right\}}_L)}$ is the collection of truth values. This is the topos classical representation of the probability of the transition $C_n\to C_{n+1\ }$. 

Now, let us define the creation operators $a^{\dagger }_{n,i}$, $i=1,2,\cdots ,m$ and $b^{\dagger }_{n,k}$, $k=1,2,\cdots ,n-\overline{\omega }$, where $n$ stands for the number of elements of $C_n$, and $i$ , and $k$ specify which element of $X_1$ and $X_2$ take part in the transition $C_n\to C_{n+1}$, respectively. It means that one can write
\begin{equation}
\left.\vert n+1,i,k\right\rangle =\sqrt{p\ }a^{\dagger }_{n,i}\left.\vert n,i,k\right\rangle +\sqrt{1-p}{\ b}^{\dagger }_{n,k}\vert\left.n,i,k\right\rangle
\end{equation}
 such that 
\begin{equation}
\left\langle n+1,i_1,k\mathrel{\left|\vphantom{n+1,i_1,k a^{\dagger }_{n,i} n,i_2,k}\right.\kern-\nulldelimiterspace}a^{\dagger }_{n,i}\mathrel{\left|\vphantom{n+1,i_1,k a^{\dagger }_{n,i} n,i_2,k}\right.\kern-\nulldelimiterspace}n,i_2,k\right\rangle ={\delta }_{i_1,i_2}, \left\langle n+1,i,k_1\mathrel{\left|\vphantom{n+1,i,k_1 b^{\dagger }_{n,k} n,i,k_2}\right.\kern-\nulldelimiterspace}b^{\dagger }_{n,k}\mathrel{\left|\vphantom{n+1,i,k_1 b^{\dagger }_{n,k} n,i,k_2}\right.\kern-\nulldelimiterspace}n,i,k_2\right\rangle =
\end{equation}
\begin{equation}
{\delta }_{k_1,k_2}, \left\langle n+1,i_1,k\mathrel{\left|\vphantom{n+1,i_1,k b^{\dagger }_{n,k} n,i_2,k}\right.\kern-\nulldelimiterspace}b^{\dagger }_{n,k}\mathrel{\left|\vphantom{n+1,i_1,k b^{\dagger }_{n,k} n,i_2,k}\right.\kern-\nulldelimiterspace}n,i_2,k\right\rangle =0,\ i_1\neq i_2
\end{equation}
and 
\begin{equation}
\left\langle n+1,i,k_1\mathrel{\left|\vphantom{n+1,i,k_1 a^{\dagger }_{n,k} n,i,k_2}\right.\kern-\nulldelimiterspace}a^{\dagger }_{n,k}\mathrel{\left|\vphantom{n+1,i,k_1 a^{\dagger }_{n,k} n,i,k_2}\right.\kern-\nulldelimiterspace}n,i,k_2\right\rangle =0,\ k_1\neq k_2
\end{equation}
or generally, if 
\begin{equation}
\left\langle n+1,i,k\mathrel{\left|\vphantom{n+1,i,k a^{\dagger }_{n,i} n,i,k}\right.\kern-\nulldelimiterspace}a^{\dagger }_{n,i}\mathrel{\left|\vphantom{n+1,i,k a^{\dagger }_{n,i} n,i,k}\right.\kern-\nulldelimiterspace}n,i,k\right\rangle =1 
\end{equation}
then $\left\langle n+1,i,k\mathrel{\left|\vphantom{n+1,i,k b^{\dagger }_{n,i} n,i,k}\right.\kern-\nulldelimiterspace}b^{\dagger }_{n,i}\mathrel{\left|\vphantom{n+1,i,k b^{\dagger }_{n,i} n,i,k}\right.\kern-\nulldelimiterspace}n,i,k\right\rangle =0$ and vice versa. Also,
\begin{equation}
\left\langle a^{\dagger }_{n,i}\mathrel{\left|\vphantom{a^{\dagger }_{n,i} a^{\dagger }_{n,j}}\right.\kern-\nulldelimiterspace}a^{\dagger }_{n,j}\right\rangle =\left\langle b^{\dagger }_{n,i}\mathrel{\left|\vphantom{b^{\dagger }_{n,i} b^{\dagger }_{n,j}}\right.\kern-\nulldelimiterspace}b^{\dagger }_{n,j}\right\rangle=0\ ,\ \ \left\langle a^{\dagger }_{n,i}\mathrel{\left|\vphantom{a^{\dagger }_{n,i} b^{\dagger }_{n,j}}\right.\kern-\nulldelimiterspace}b^{\dagger }_{n,j}\right\rangle =0
\end{equation}
 because at each step, the new element attaches to only one element of $C_n$ .

By considering, $a^{\dagger }_{n,i}$ and ${\ b}^{\dagger }_{n,k}$ as pairwise orthogonal projection operators, then $\mathcal{V}\left(\mathcal{H}\right)=\{a^{\dagger }_{n,1},\ a^{\dagger }_{n,2},\cdots ,a^{\dagger }_{n,m},\ b^{\dagger }_{n,1},\ b^{\dagger }_{n,2},\ \cdots ,b^{\dagger }_{n,n-\overline{\omega }}\}$, $\mathrm{\Gamma }\left({\underline{\mathrm{\Omega }}}^{sh{(\mathcal{V}(\mathcal{K})\times (0,1)}_L)}\right)=\mathrm{\Gamma }\left({\underline{\mathrm{\Omega }}}^{sh{(S\times \{P\}}_L)}\right)$, and ${\epsilon }^{\mu }:Sh({\left\{P\right\}}_L)(\underline{X})\to \mathrm{\Gamma }{\underline{\mathrm{\Omega }}}^{Sh(\mathcal{V}\left(\mathcal{H}\right)\times {\left\{0,1\right\}}_L)}$, where ${\left\{P\right\}}_L$ is the topological space, $Sh({\left\{0,1\right\}}_L)(\underline{X})$ is constant sheaf and $\mathrm{\Gamma }{\underline{\mathrm{\Omega }}}^{Sh({\mathcal{V}\left(\mathcal{H}\right)\times \left\{0,1\right\}}_L)}$ is the collection of truth values. This is the topos quantum representation of the probability of the transition $C_n\to C_{n+1\ }$.

Since, the set $P=\{a^{\dagger }_{n,1},\ a^{\dagger }_{n,2},\cdots ,a^{\dagger }_{n,m},\ b^{\dagger }_{n,1},\ b^{\dagger }_{n,2},\ \cdots ,b^{\dagger }_{n,n-\overline{\omega }}\}$ is pairwise orthogonal then $({\mu }_{\vert V}\left(a^{\dagger }_{n,1}\right),\ {\mu }_{\vert V}\left(a^{\dagger }_{n,2}\right),\ \cdots ,\ {\mu }_{\vert V}\left(b^{\dagger }_{n,n-\overline{\omega }}\right))$ is the probability distribution and the contextual entropy
\begin{equation}
 E{(\mu )}_{\vert V}=Sh\left({\mu }_{\vert V}\left(a^{\dagger }_{n,1}\right),\ {\mu }_{\vert V}\left(a^{\dagger }_{n,2}\right),\ \cdots ,\ {\mu }_{\vert V}\left(b^{\dagger }_{n,n-\overline{\omega }}\right)\right)=
\end{equation}
\begin{equation}
-\sum^{m+n-\overline{\omega }}_{i=1}{{\mu }_{\vert V}\left(P_i\right)\mathrm{ln}\mathrm{}({\mu }_{\vert V}\left(P_i\right))}
\end{equation}
is the Shannon entropy. Also, we mentioned above, the contextual entropy $E{({\mu }_{\rho })}_{V_{\rho }}$ is von Neumann entropy, if the eigenvalues of $\rho $ is written as ${\lambda }_i=Tr(\rho P_i)$. Now, by growing the network of space-time, the number of elements of the set $P$ increases, and inconsequence, $E{(\mu )}_{\vert V}$ and $E{({\mu }_{\rho })}_{V_{\rho }}$ increase. It means that the network will be more stable than before. Then, why we see universe inflation because its entropy increases due to the increment of the space-time events [25].

\section{Vibration of Space-Time Network}

In solid-state physics, it is assumed that the atoms of crystal vibrate around their equilibrium position else at zero Kelvin. One can assign a pseudo-particle, called acoustic phonon, to the collection of vibration of similar atoms (such as Gallium-Gallium and Arsenide-Arsenide in Gallium Arsenide crystal) and an optical phonon to the collection of vibration of different atoms (such as Gallium-Arsenide in Gallium Arsenide crystal) in the crystal. Different acoustic and optical phonons with different energies exist the crystal due to the different vibration modes [64]. An emission spectrum can be assigned to the solid whose temperature is higher than the environment. The Krichhoff's law relates the emission to the emission of black body at the same temperature [65]. If the true reflectivity is the fraction of incident radiation that is reflected from the first surface and the true transmittance is the fraction of light entering the solid that reaches the second surface, it can be shown that, there is some expressions for the apparent reflectivity, the apparent transmissivity, and the emissivity in terms of the true reflectivity and the true transmittance [65]. It can be shown that the direct measurement of the emittance offers a distinct advantage over the reflectance and transmittance measurements [65]. The determination of the absorption coefficient of some materials by transmission method would have required very thick samples in the regions where the absorption coefficient is as low as $0.2\ {Cm}^{-1}$ [65]. Therefore, the emission technique seems eminently suitable for the study of absorptance in the overtone and the combination regions [65]. Also, the inelastic scattering of photons by phonons is known as the Raman effect in solids. The Stokes (energy lost) or anti-Stokes (energy gained) effects by the photon in such a process is accomplished by the creation or annihilation of a phonon. Brillouin scattering is a special case of Raman scattering, involving the low-frequency acoustic phonons [65]. Therefore, infrared and Raman spectroscopies are practical methods for studying the structure of crystalline solids [64, 65]. It should note that a crystalline solid is an ordered network i.e., one can construct the entire crystal by repeating its unit cell.

Generally, a network is not crystalline. A question can be asked: whether one can assign phonons to the vibrations of the network?  Let us to build a Watta-Strogatz (WS) network [66]. First, we consider an one dimensional network with only local connections of range $r$ which is constructed under the periodic conditions [66]. Next, each local edge (or link) is visited once, and with rewiring probability $P$ removed and reconnected to a randomly chosen node [66]. The entire network is swept and the number of shortcuts in the system of size $N$ is given by $N\ P\ r$ [66]. Now, an atom is put on each node whereas an edge connecting two nodes represents the coupling between two atoms [66]. If it is assumed that all atoms are identical, each having mass M and moving only along the direction of the chain, the equation of motion for the $l$-th atom at the position $x_l$ reads $M{\ddot{x}}_l=C\sum_{m\in {\mathrm{\Lambda }}_l}{(x_m-x_l)}$, where ${\mathrm{\Lambda }}_l$ stands for the set of nodes connected to the $l$-th node and the interaction strength $C$ is assumed to be the same for any pairs of the interacting atoms [66]. In the presence of shortcuts ($P\neq 0$), the equation of motion in the matrix form is
\begin{equation}
{\omega }^2\boldsymbol{X}=\boldsymbol{DX}
\end{equation}
where $\boldsymbol{X}$\textbf{ }is the\textbf{ }$N$-dimensional column vector with the component ${\overline{x}}_i(i=1,\cdots ,N)$ and $\boldsymbol{D}$ is $N\times N$ dynamic matrix with the elements
\begin{equation}
D_{ij}=\{\left|{\mathrm{\Lambda }}_i\right|\ for\ i=j;\ -1\ for\ j\in {\mathrm{\Lambda }}_i;0\ for\ otherwise\}
\end{equation}
 where ${\mathrm{\Lambda }}_i$ denotes the number of nodes connected directly to the $i$-th node via local links or shortcuts [66]. It has been shown that one can assign the quants of vibration, called netons, to the vibration modes of the network [66]. The density of neton levels, obtained numerically, reveals that unlike a local regular lattice, the system develops a gap of finite width, manifesting extreme rigidity of the network structure at low energies [66].  Also, one can consider a complex network such that the balls of mass $m$ are placed on the nodes and links are springs with the springs constant $m{\omega }^2$ which connecting two balls [67]. It is assumed that the ball-spring network is submerged into a thermal bath at the temperature $T$ and the balls in the complex network oscillate under thermal disturbance [67]. It can be shown that the Hamiltonian of the classical system can be written as
\begin{equation}
H_L=\sum_i{\frac{p^2_i}{2m}+\frac{m{\omega }^2}{2}}\sum_{i,j}{x_iL_{ij}x_j}
\end{equation}
where $L_{ij}$ denotes an element of the network Laplacian. The network Laplacian is given by $L=D-A$, where $D$ is a diagonal matrix with $D_{ii}=k_i$ and $A$ is adjacency matrix with elements [67]
\begin{equation}
A_{ij}=\{1\ for\ j=i+1;0\ for\ j\neq i;\ -1\ for\ j=i-1\}
\end{equation}
Also, the Hamiltonian of the quantum system can be written as $H_A=\sum_{\mu }{H_{\mu }}$ where
\begin{equation}
H_{\mu }=\hslash \mathrm{\Omega }\left[1+\frac{{\omega }^2}{2{\mathrm{\Omega }}^2}({\lambda }_{\mu }-K)\right]\left(b^{\dagger }_{\mu }b_{\mu }+\frac{1}{2}\right)
\end{equation}
 Here, the constant $K\ge {max}_ik_i$ where $k_i$ is the number of links that are connected to the node $i$ (the degree of the node $i$), $\mathrm{\Omega }=\sqrt{K/m\omega }$, and ${\lambda }_{\mu }$ is the eigenvalues of $\mathrm{\Lambda }=O(KI-A)O^T$ where $I$ and $A$ are unit matrix and adjacency matrix, respectively and $O$ is an orthogonal matrix. Also, $b_{\mu }=\sum_i{a_i{\left(O^T\right)}_{i\mu }}$ and $b^{\dagger }_{\mu }=\sum_i{a^{\dagger }_i{\left(O^T\right)}_{i\mu }}$ where, $a_i=\frac{1}{\sqrt{2\hslash }}(x_i\sqrt{\mathrm{m}\mathrm{\Omega }}+\frac{i}{\sqrt{m\mathrm{\Omega }}}p_i)$ and $a^{\dagger }_i=\frac{1}{\sqrt{2\hslash }}(x_i\sqrt{\mathrm{m}\mathrm{\Omega }}-\frac{i}{\sqrt{m\mathrm{\Omega }}}p_i)$ which $x_i{-x}_j$ is the extension or the contraction of the spring connecting the nodes $i$ and $j$ and $p_i$ stands of momentum of node $i$ [67]. 

As we mentioned above, when we deal with the space-time network, we consider the action $S=\frac{-1}{8\pi G}\int{\left(\mathrm{\Lambda }\right)\sqrt{-g}d^4x}$  which does not include the matter or gravitational waves [53]. Then, how one can discuss about the vibration of the space-time network? There are two scenarios. First, he/she generalizes the equation ${\omega }^2\boldsymbol{X}=\boldsymbol{DX}$, where $\omega $ stands for resulted interaction and causal relationship between nodes (locales) through the links which has led to the formation of a relaxed (optimized) space-time network at every moment (classical snapshot of space-time), although both temperature and pressure were high in the early moments of the formation of the universe. Therefore, it is expected that the netons can be attributed to the vibration modes of the space-time network with a gap of finite width, manifesting extreme rigidity of the space-time network structure at low energies. The other solution is that, he/she generalizes the Hamiltonian $H_{\mu }=\hslash \mathrm{\Omega }\left[1+\frac{{\omega }^2}{2{\mathrm{\Omega }}^2}({\lambda }_{\mu }-K)\right]\left(b^{\dagger }_{\mu }b_{\mu }+\frac{1}{2}\right)$ where $\omega $ is similar to the previous case and $K$ is some unknown constant. But, he/she can not specify which of the homomorphic space-time networks can be attributed to the probable spectrum of netons, even if in future, the experimental data will be found in the cosmic background radiation. It should be noted that in topos representation, the space-time network is mapped to the topos quantum manifold with two topologies and a locale is assigned to each node. Therefore, $x_i{-x}_j$ is the extension or the contraction of distance between two neighborhood locales which are submerged into a thermal bath at the temperature $T$ and the locales in the complex network oscillate under thermal disturbance. In second scenario, he/she can assume that the submerged locales into a thermal bath have Brownian motions similar to a free particle in a fluid which is adequately described by the Wiener process $w(t)$ [68]. The Wiener process is characterized by four properties: (i) $w_0=0$, (ii) $w$ had independent increments: for every $t>0$, the future increments $w_{t+u}-w_t\ge 0,\ u\ge 0$ are independent of the past values $w_s,\ s<t$, (iii) $w$ has Gaussian increments: $w_{t+u}-w_t$ is normally distributed with mean 0 and variance $u$, $w_{t+u}-w_t\sim N(0,u)$, and (iv) $w$ has continuous paths: $w_t$ is continuous in $t$ [69]. If $E$ and $v$ stand for the expectation (average, $\hslash /2m$) and the diffusion coefficient ($\frac{kT}{m\beta },$ which $m\beta $ is the friction coefficient), respectively, then $E\ dw_i\left(t\right)\ {dw}_j\left(t\right)=2v{\delta }_{ij}\ dt$ [68]. It can be shown that, every particle of mass $m$ which is subjected to a Brownian motion with diffusion coefficient $\hslash /2m$ and zero friction and is under influence of an external electromagnetic field obeys the Schrodinger equation with entirely classical interpretation [68]. Also, it can be shown that the quantum theory can be arisen from quantum gravity when a graph $\mathrm{\Gamma }$ with a finite set of $N$ nodes is embed in ${\mathcal{R}}^3$ and is subject to some microscopic rules of evolution i.e., the microscopic model is a spin foam with a single in and out graph and presumably a single interpolating history [70]. The stochastic part of the evolution of the node coordinates could be due either to the existence of hot regions in the embedded model or coarse-grained description of several of the underlying nodes [70]. Now, if one assumes that the primitive creatures of the creation, due to being in a thermal bath or coarse-graining process, have a Brownian movement that is explained by the Wiener process, then he/she can find that the equation describing their movements is
\begin{equation}
\left[-\frac{{\hslash }^2}{2m}{\mathrm{\nabla }}^2+V-E\right]\psi =0 
\end{equation}
at each classical snapshot [70] i.e., the equation of harmonic oscillator, $H_{\mu }=\hslash \mathrm{\Omega }\left[1+\frac{{\omega }^2}{2{\mathrm{\Omega }}^2}({\lambda }_{\mu }-K)\right]\left(b^{\dagger }_{\mu }b_{\mu }+\frac{1}{2}\right)$, which is satisfied at each classical snapshot which $\omega $ stands for resulted interaction and causal relationship between nodes (locales) through the links which has led to the formation of a relaxed (optimized) space-time network at every moment (classical snapshot of space-time) and $K$ is some unknown constant. Therefore, we expect to observe the spectrum of netons (similar to infrared and Raman spectrum) in the cosmic background radiation [26, 71]. If the total Hamiltonian of primitive creatures includes some other terms such as $\sum_i{\sum_j{J_{ij}{\hat{S}}_{1i}{\hat{S}}_{2j}}}$ (spin-interaction), $\sum_i{\sum_j{Q_{ij}{\hat{I}}_{1i}{\hat{I}}_{2j}}}$ (charge-interaction) and so on, it is expected that their signature will be found in the cosmic background radiation  [26, 71]. In this case, we should consider a new action (not $S=\frac{-1}{8\pi G}\int{\left(\mathrm{\Lambda }\right)\sqrt{-g}d^4x}$) which includes the interaction terms [53]. Therefore, our previous discussion about inflation of universe should be revised and a new one should be provided in future. 

\section{Non-commutative Geometry}

The simplest case of noncommutative coordinate product is given by $\left[x^{\mu },x^v\right]=i{\mathrm{\Theta }}^{\mu v}$ which the antisymmetric tensor ${\mathrm{\Theta }}^{\mu v}$ is a $c$-number, with $\mu ,\ v=0,\cdots ,n$, where $n+1$ is the dimension of the space-time, and accounts for the degree of quantum fuzziness of space-time [72]. Also, in the Lie-algebra and the coordinate-dependent ($q$-deformed) formulation, the noncommutativity of the coordinates could take place [73]. In the other words, the associative algebra structure $A_x$, which defines a noncommutative space, can be defined in terms of a set of generators ${\hat{x}}^i$ and relations $\mathcal{R}$ [73]. These are of the form of a canonical structure 
\begin{equation}
\left[{\hat{x}}^i,\ {\hat{x}}^j\right]=i{\theta }^{ij},\ {\theta }^{ij}\in \mathbb{C}
\end{equation}
a Lie-algebra structure
\begin{equation}
\left[{\hat{x}}^i,\ {\hat{x}}^j\right]=iC^{ij}_k{\hat{x}}^k,\ C^{ij}_k\in \mathbb{C}
\end{equation}
and a quantum space structure 
\begin{equation}
{\hat{x}}^i{\hat{x}}^j=q^{-1}{\widehat{\mathcal{R}}}^{ij}_{kl}{\hat{x}}^k{\hat{x}}^l,\ {\widehat{\mathcal{R}}}^{ij}_{kl}\in \mathbb{C}
\end{equation}
 where $i=1,\cdots ,N$ [73]. 

But, the above noncommutative relationships can be interpreted from the perspective of measurement theory. Let us to expose a particle to light (gamma) ray radiation using the lens of a (Heisenberg) microscope. If $\lambda $ is the wave length of the radiation that enters the lens $L$, and $\theta $ is the half angle subtended at the particle $P$ by the lens, the best resolving power of the lens $L$ is known to provide an accuracy $\Delta x=\frac{\lambda }{sin\theta }$ in a position determination [74]. Now, if $\left[x^{\mu },x^v\right]=i{\mathrm{\Theta }}^{\mu v}$ and $\Delta x^{\mu }=\frac{\lambda }{sin\theta }$ then $\Delta x^v\propto \frac{sin\theta }{\lambda }$. It means that increment in the measurement accuracy of $\Delta x^{\mu }$ Leading to  the decrement in the measurement accuracy of $\Delta x^v$ and vice versa.

However, it is well known that, If the state $\psi $ is not an eigenstate of operator $\hat{A}$, then one would get a certain range $\mathrm{\Delta }$ of values for observable $A$, where $\mathrm{\Delta }\in sp(\hat{A})$,and $sp$ stands for spectrum. In topos language, it means that $\hat{A}$ gets approximated both from above (${\check{\delta }}^o{(\hat{A})}_V=v$) and from below (${\check{\delta }}^i{(\hat{A})}_V=\mu $) [26, 45].  Such an approximation gets coarser as $W\subseteq V\in \mathcal{V}(\mathcal{H})$ (gets smaller), which basically means that $W$ contains less and less projections, i.e., less and less information [26, 45]. Therefore, as we mentioned before, if the space-time network includes $N$-event (point) we can attribute to each point a locale ${{\underline{R}}^{\leftrightarrow }}_i(\mu ,\ v),i=1,2,\cdots ,N$, at each classical snapshot, $j$. Therefore, for leading to the increment in the measurement accuracy of $\Delta x^{\mu }$, at each classical snapshot, $j$, we need to get more information, which is equivalent to using bigger algebra than the previous measurement i.e., we need to get bigger  ${{\underline{R}}^{\leftrightarrow }}_i(\mu ,\ v)$. \textbf{ }However, increasing ${{\underline{R}}^{\leftrightarrow }}_i(\mu ,\ v)$ means that the space occupied by point $i$ has increased, and therefore, if it is increased to its maximum limit to obtain sufficient accuracy (information), then the recognition of other points ($l$), especially neighboring points, from ${{\underline{R}}^{\leftrightarrow }}_l(\mu ,\ v)$ will be accompanied by a maximum error.  Perhaps, from the measurement process point of view, this phenomenon can be described as such that if we look at space ${{\underline{R}}^{\leftrightarrow }}_i(\mu ,\ v)$ with a Heisenberg microscope, it should be in such a way that we can sweep the maximum space to find the point $i$, and therefore it will not be possible to observe other points, ${{\underline{R}}^{\leftrightarrow }}_l(\mu ,\ v)$ with enough accuracy. Therefore, it can be said that the noncommutative geometry is rooted in the fact that the creatures of the early moments of creation occupy some space of space-time i.e., locale instead of a pint. 

\section{Conclusion}

\noindent Among the various existing theories, we have shown that how the concept of the space-time network has entered the physics of quantum gravity by reviewing the theories of loop quantum gravity causal sets. Assuming that the first creatures of creation create a network, it has been shown that how the network can be mapped to a topos discrete quantum manifold which has been equipped with two algebras. One of these algebras is classical and equipped with discrete calculus, and the other is Alexandrov's algebra and belongs to the topos theory. We have assigned a locale to each nodes of the space-time network and shown that in general, invariance under Lorentz transformations is no longer true. Using the implicit concept in the locale, it was shown that the cosmological constant is non-zero and can be proportional to the second power of the Hubble radius. By considering a population (set), including newly born timid children and non-timid children who survive until the birth of the new network, we have shown that the entropy of the aforementioned set is quantized and increases as the set grows. In consequence, the inflation of the world is expected phenomenon. Also, we have shown that how world inflation can be described based on the concept of truth object and truth value belong to the topos theory. Basically, this space-time network is momentarily stable and relaxed (optimized) during the inflation of the universe, although the temperature and pressure are both high. We have shown that the quanta of vibrations, called netons, can be attributed to the vibration of space-time network, and it is expected that we observe them in the future experiments related to cosmic background radiation. Finally, it is shown that the root of noncommutative geometry is in attributing the locale to the nodes of the space-time network instead of a point. This theory, which is quantum-relativistic from the beginning, has not the problem of a point particle, the concept of probability is an emergent concept, it does not include the problem of the measurement theory, it is not a universal Lorentz invariant, its cosmological constant is non-zero and it can be proportional to the power the second power of the Hubble radius, it describes the inflation of the universe and it has a non-commutative geometry, is called the many-node theory.

\textbf{References:}

\noindent [1] D. Rickles, S. French, and J. Saatsi, ``The structural Foundations of Quantum Gravity'' (Oxford University Press, 2006).

\noindent [2] B. P. Abbott et al., ``Observation of Gravitational Waves from a Binary Black Hole Merger'', Phys. Rev. Lett. \textbf{116}, 061102 (2016).

\noindent [3] Abhay Ashtekar, ``New Variables for Classical and Quantum Gravity'', Phys. Rev. Lett. \textbf{57}, 18, (1986).

\noindent [4] R. Gambini and J. Pullin, ``A first Course in Loop Quantum Gravity'' (Oxford University Press, 2011).

\noindent [5] Robert M. Wald, ``General Relativity'' (The University of Chicago Press, 1984).

\noindent [6] Abhay Ashtekar, ``New Hamiltonian formulation of general relativity'', Phys. Rev. D, \textbf{36}, 6 (1987).

\noindent [7] A. Ashtekar, J. D. Romano, and R. S. Tate, ``New variables for gravity: Inclusion of matter'', Phys. Rev. D \textbf{40}, 8 (1989).

\noindent 
\paragraph{[8] R. Gambini and A. Trias, ``Second quantization of the free electromagnetic field as quantum mechanics in the loop space ``, Phys. Rev. D 22, 1380 (1980).}

\noindent 
\paragraph{[9] R. Gambini and A. Trias, ``Geometrical origin of gauge theories ``, Phys. Rev. D 23, 553 (1981).}

\noindent 
\paragraph{[10] R. Gambini and A. Trias, ``Gauge dynamics in the C-representation ``, Nuc. Phys. B 278, 436 (1986).}

\noindent 
\paragraph{[11] C. Rovelli and L. Smolin, ``Knot Theory and Quantum Gravity ``, Phys. Rev. Lett. 61, 1155 (1988).}

\noindent [12] R. Giles, ``Reconstruction of gauge potentials from Wilson loops'', Phys. Rev. D \textbf{24}, 2160 (1981).

\noindent 
\section{[13] R. Gambini and J. Pullin, ``Loops, Knots, Gauge Theories and Quantum Gravity'' (Cambridge University Press, 1996).}

\noindent [14] C. Rovelli, ``Quantum Garvity'' (Cambridge University Press, 2010).

\noindent [15] C. Rovelli and F. Vidotto, ``Covariant Loop Quantum Gravity'' (Cambridge University Press, 2014).

\noindent [16] Hermann Weyl, ``Space-Time-Matter'' (Dover Publications, Inc. 1950).

\noindent [17] E. C. Zeeman, ``Causality Implies the Lorentz Group'', J. Math. Phys. \textbf{5}, 4 (1964).

\noindent [18] S. W. Hawking, A. R. King, and P. J. McCarthy, ``A new topology for curved space-time which incorporates the causal, differential, and conformal structures'' J. Math. Phys. \textbf{17}, 174 (1976).

\noindent [19] J. Myrheim, ``Statistical Geometry'', CERN-Geneva Ref.TH.2538-CERN (1978).

\noindent [20] G. `t Hooft, ``Quantum gravity: a fundamental problem and some radical ideas'' (in M. Levy and S. Deser, eds., Recent Developments in Gravitation, Cargese 1978, Plenum 1979). 

\noindent [21] Luca Bombelli, ``PhD Thesis: Space-Time as a Causal Set'' (Syracuse University, 1983).

\noindent [22] L. Bombelli, J. Lee, D. Meyer, and R. D. Sorkin, ``Space-Time as a Causal Set'', Phys. Rev. Lett. \textbf{59}, 5 (1987).

\noindent [23] D. P. Rideout and R. D. Sorkin, ``Classical Sequential Growth Dynamics foe Causal Sets'', Phys. Rev. D \textbf{61}, 024002 (1999).

\noindent [24] Sumati Surya, ``The Causal Set Approach to Quantum Gravity'' Living Reviews in Relativity \textbf{22}, 5 (2019).

\noindent [25] Hamidreza Simchi, ``Statistical Geometry Representation of Spacetime'' (Chapter 8 of A. Mohammed ed. New Frontiers in Physical Science Research Vol.9, B P International, 2023, DOI: 10.9734/bpi/nfpsr/v9/9579F).

\noindent [26] Hamidreza Simchi, ``General Formulation of Topos May-Node Theory'' (Chapter 7 of A. Mohammed ed. New Frontiers in Physical Science Research Vol.9, B P International, 2023, DOI: 10.9734/bpi/nfpsr/v9/9580F).

\noindent [27] J. Reiterman, V. Rodl, and E. Sinajova, ``Geometrical Embedding of Graphs'', Discrete Math. \textbf{74}, 291 (1989).

\noindent [28] Mikio Nakahara, ``Geometry, Topology, and Physics'' (Institute of Physics Ltd, 2003).

\noindent [29] A. Dimakis and F. Muller-Hoissen, ``Discrete Differential Calculus: Graphs, Topologies, and Gauge Theory'', J. Math. Phys. \textbf{35}, 6703 (1994).

\noindent [30] A. Dimakis, F. Muller-Hoissen, and F. Vanderseypen, ``Discrete Differential Manifolds and Dynamics on Networks'', J. Math. Phys. \textbf{36}, 3771 (1995).

\noindent [31] L. M. Chen, ``Digital and Discrete Geometry'' (Springer, 2014). 

\noindent [32] T. D. Lee, ``Can Time be a Discrete Dynamical Variable?'', Phys. Lett. B \textbf{122}, 217 (1983).

\noindent [33] C. J. Isham, ``An introduction to general topology and quantum topology'', (in Proceedings of the Advanced Summer Institute on Physics, Geometry and Topology, 1989).

\noindent [34] C. J. Isham, ``Quantum topology and quantization on the lattice of topologies'', Class. Quantum Grav. \textbf{6}, 1509 (1989).

\noindent [35] R. D. Sorkin, ``Finitary Substitute for Continuous Topology'', Inter. J. Theoretical Phys. \textbf{30}, 923 (1991).

\noindent [36] A. P. Balachandran, G. Bimonte, E. Ercolessi, and P. T. Sobrinho, ``Finite approximations to quantum physics: Quantum points and their bundles'', Nuc. Phys. B \textbf{418}, 477 (1994).

\noindent [37] Manuel Hohmann, ``Quantum Manifold'', AIP Conf. Proce. \textbf{1424}, 149 (2012) (arXiv:0809.3111~[math-ph])

\noindent [38] Hamidreza Simchi, ``The Concept of Time: Causality, Precedence, and Space Time'' (Chapter 7 of A. Mohammed ed. New Frontiers in Physical Science Research Vol.9, B P International, 2023, DOI: 10.9734/bpi/nfpsr/v9/9578F).

\noindent [39] Thomas F. Jordan, ``Quantum Mechanics in Simple Matrix Form'' (John Wiley \& Sons, 1936). 

\noindent [40] H. S. Green and M. Born, ``Matrix Mechanics'' (P. Noordhoff LTD, 1965).

\noindent [41] Alexander Komech, ``Quantum Mechanics: Genesis and Achievements'' (Springer, 2013).

\noindent [42] Steven Weinberg, ``The troubles with Quantum Mechanics'', NYREV, Inc. January 19 (2017).

\noindent [43] Lee Smolin, ``Trouble with Physics'', (Houghton Mifflin Harcourt, 2006).

\noindent [44] Hugh Everett, ``The Theory of the Universal Wave Function'', (PhD Thesis, Princeton. Reprinted in De Witt and Graham, 1973).

\noindent [45] Cecilia Flori, ``A First Course in Topos Quantum Theory'', (Springer, 2013).

\noindent [46] Cecilia Flori, ``A Second Course in Topos Quantum Theory'', (Springer, 2018).

\noindent [47] Cecilia Flori, ``Group Action in Topos Quantum Physics'', arXiv: 1110.1650 [quant-ph].

\noindent [48] Hamidreza Simchi, ``A Grand Unified Reaction Platform: Concept of Time Approach'' (Chapter 7 of A. Mohammed ed. New Frontiers in Physical Science Research Vol.9, B P International, 2023, DOI: 10.9734/bpi/nfpsr/v9/9570F).

\noindent [49] J. Ambjorn, J. Jurkiewicz, and R. Loll, ``The Universe from Scratch'', Contemp. Phys. \textbf{47}, 103 (2006).

\noindent [50] C. J. Isham and J. Butterfield, ``Some possible roles for topos theory in quantum theory and quantum gravity'', Found. Phys. \textbf{30}, 1707 (2000).

\noindent [51] Sarita Dalya Rosenstock, ``A Categorial Consideration of Physical Formalisms'' (PhD Thesis, University of California, Irvine, 2019).

\noindent 

\noindent [52] L. M. Gaio and B. F. Rizzuti, ``A categorical View on the Principle of Relativity'', arXiv:2205.15408v1 [math-ph].

\noindent [53] L. D. Landau and E. M. Lfshitz, ``The Classical Theory of Fields'' (Pergamon, Oxford, 1951).

\noindent [54] M. Ahmed, S. Dodelson, Patrick B. Green, and R. Sorkin, ``Everpresent $\mathrm{\Lambda }$'', Phys. Rev. D \textbf{69}, 103523 (2004).

\noindent [55] Nosiphiwo Zwane, ``Cosmological Tests of Causal Set Phenomenology'' (PhD Thesis, University of Waterloo, 2017).

\noindent [56] F. Dowker and S. Zalel, ``Evolution of Universe in Casual Set Cosmology'', C. R. Physique \textbf{18}, 246 (2017). 

\noindent [57] David P. Rideout, ``Dynamics of Causal Sets'' (PhD Thesis, Georgia Institute of Technology, 1995).

\noindent [58] D. P. Rideout and R. D. Sorkin, ``Classical sequential dynamics for causal sets'', Phys. Rev. D \textbf{61}, 024002 (1999). 

\noindent [59] F. Dowker and S. Zalel, ``Evolution of universe in causal set cosmology'', Comptes Rendus Physique \textbf{18}, 3 (2017).

\noindent [60] R. D. Sorkin, ``Indication of causal set cosmology'', Int. J. Theor. Phys. \textbf{39}, 1731 (2000).

\noindent [61] Djamel Dou, ``Causal sets, a possible interpretation for the black hole entropy and related topics'' (PhD Thesis, SISSA, Trieste, 1999). 

\noindent [62] L. Demetrius, ``Natural selection and age-structured populations'', Genetic, \textbf{79}, 535 (1975).

\noindent [63] C. M. Constantin and A. Doring, ``A Topos Theoretic Notion of Entropy'', arXiv:2006.03139v1 [math.CT] (2020).

\noindent [64] J. M. Ziman, ``Electrons and Phonons: The Theory of Transport Phenomena in Solids'' (Oxford, 1960).

\noindent [65] Shashanka S. Mitra, ``Infrared and Raman Spectra Due to Lattice Vibrations'' (Chapter 14 of S. Nudelman et al. (eds.), Optical Properties of Solids, Springer Science - Business Media New York 1969).

\noindent [66] B. J. Kim, H. Hong, and M. Y. Choi, ``Netons: vibrations of complex networks'', J. Phys. A: Math. Gen. \textbf{36}, 6329 (2003).

\noindent [67] E. Estrada, N. Hatano, and M. Benzi, ``Yhe physics of communicability in complex networks'', Phys. Reports \textbf{514}, 89 (2012).

\noindent [68] Edward Nelson, ``Derivation of the Schrodinger Equation from Newtonian Mechanics'', Phys. Rev. \textbf{150}, 4 (1966).

\noindent [69] Dominic Breit, ``Existence Theory for Generalized Newtonian Fluids'' (Elsevier, Academic Press, 2017).

\noindent [70] F. Markopoulou and L. Smolin, ``Quantum theory from quantum gravity'', Phys. Rev. D \textbf{70}, 124029 (2004). 

\noindent [71] J. Bluemer,~R. Engel,~J.R. Hoerandel, ``Cosmic Rays from the Knee to the Highest Energies'' Prog. Part. Nucl. Phys. \textbf{63}, 293 (2009).

\noindent [72] T. Kanazawa, G. Lambiase, G. Vilasi, and A. Yoshioka, ``Noncommutative Schwarzchild geometry and generalized uncertainty principle'', Eur. Phys. J. C \textbf{79}, 95 (2019).

\noindent [73] J. Madore, S. Schraml, P. Schupp, and J. Wess, ``Gauge theory on noncommutative space'', Eur. Phys. J. C \textbf{16}, 161 (2000).

\noindent [74] Leonard L. Schiff, ``Quantum Mechanics'' (McGraw-Hill Company, 1968).

\noindent

\noindent

\end{document}